\documentclass[10pt,a4paper]{article}
\PassOptionsToPackage{table}{xcolor}
\usepackage{amsmath, amssymb, graphicx}
\usepackage{natbib}
\usepackage{latexsym} 
\usepackage{caption} 
\usepackage[utf8]{inputenc}
\usepackage{bm}
\usepackage{tikz}
\usepackage{appendix}
\usepackage{multirow}
\usepackage{url}
\usepackage{longtable}

\usepackage[mode=buildnew]{standalone} 
\usepackage{kbordermatrix,blkarray} 
\usepackage{siunitx}

\graphicspath{{figures/}{../figures/}{./}}

\newcommand{\CO}{$\text{CO}_2$ }
\DeclareMathOperator*{\argmax}{arg\,max}

\begin{document}
\author{Unn Dahl\'en, Johan Lindstr{\"o}m and Marko Scholze}
\title{Spatio-Temporal Reconstructions of Global CO2-Fluxes using Gaussian Markov Random Fields}

\maketitle

\begin{center}\bf{Abstract}\end{center}
Atmospheric inverse modelling is a method for reconstructing historical fluxes of green-house gas between land and atmosphere, using observed atmospheric concentrations and an atmospheric tracer transport model.
The small number of observed atmospheric concentrations in relation to the number of unknown flux
components makes the inverse problem ill-conditioned, and assumptions
on the fluxes are needed to constrain the solution. A common practise
is to model the fluxes using latent Gaussian fields with a mean
structure based on estimated fluxes from combinations of process 
modelling (natural fluxes) and statistical bookkeeping (anthropogenic
emissions). Here, we reconstruct global \CO 
flux fields by modelling fluxes using Gaussian Markov Random Fields (GMRF), resulting in a flexible and computational beneficial model with a Mat\'ern-like spatial covariance, and a temporal covariance defined through an auto-regressive model with seasonal dependence. 
 
In contrast to previous inversions, the flux is defined on a spatially continuous
domain, and the traditionally discrete flux representation is replaced
by integrated fluxes at the resolution specified by the transport
model. This formulation removes aggregation errors in the flux
covariance, due to the traditional representation of area integrals by
fluxes at discrete points, and provides a model closer resembling real-life space-time continuous fluxes.

\begin{flushleft}
\end{flushleft}

\section{Introduction}
The steady increase of atmospheric greenhouse gases since the 18$^{\text{th}}$
century industrial revolution, has been attributed to anthropogenic
emissions from mainly fossil fuel burning and land-use changes
\citep{lequere18}. In order to limit future global warming associated to increasing
greenhouse gas (GHG) concentrations in the atmopshere, we need to better
understand the sources and sinks of the GHG fluxes caused by
interactions of climate, ecosystems and human 
activities. 

Because direct observations of trace gas fluxes are limited both spatially and in
time, the spatio-temporal flux patterns are inferred from mathematical
models. Besides bottom up modelling based on prognostic models
including process understanding,
a top down approach based on observations of atmospheric GHG
concentrations is commonly used. Since \CO is a persistent GHG, the atmosphere itself contains 
information regarding past fluxes. In combination with models for
atmospheric transport, the atmospheric concentrations can be used to
reconstruct past fluxes; this is the main idea behind atmospheric
inverse modelling \citep{rayner1999,enting2002}. 

The atmospheric concentrations are sampled at
several sites across the globe. The high ratio of unknown flux components to observations makes the inversion problem ill-conditioned. 
Therefore, atmospheric inversion studies have adapted
a Bayesian approach by assigning a (Gaussian) prior to the fluxes. The
expectation (or prior fluxes) in this prior field combines estimates of natural
biosphere fluxes from process-based Dynamical Global Vegetation Model, such as the 
Biosphere Energy Transfer and Hydrology (BETHY) scheme \citep{knorr2000} 
with  measurements of ocean fluxes \citep[][e.g.]{takahashi2002}. Fossil fuel emissions \citep[][e.g.]{boden2011} are usually taken as
known and are pre-subtracted from the observations.

Many inversion set-ups have been based
on low resolution surface fluxes (e.g.\ at a continental
scale), wherein the fluxes have been assumed to be piecewise constant
\citep{gurney2003,gurney2004,law2003,baker2006}.
This "big region" approach results in fluxes that are constrained by
the observations. However, with a transport model of higher resolution
than that of the fluxes, this approach is prone to potential aggregation errors, as
well as limited resolution of the estimated posterior fluxes. To overcome these issues, studies like \citet{houweling1999, kaminski1999b,
  rodenbeck2003, michalak2004,peters2005,zupanski2007,mueller2008},
resolved the fluxes at a higher  resolution, comparable with that of
the transport models. Due to the larger number of unknown fluxes,
these "grid-scale" studies are even more reliant on prior assumptions
on the fluxes. Most of the studies assume a Gaussian prior with a priori known dependence structure for the fluxes; \citet{michalak2004} introduced a geostatistical approach, where unknown parameters in both a regression model for the prior expected fluxes and the spatial dependence structure were inferred from data, thus reducing the extent of prior assumptions. Parallel to the standard inversion techniques,
recursive estimation methods, such as Ensemble Kalman Filters (EKF),
have been investigated as potential tools for atmospheric inversions
\citep{peters2005,zupanski2007}. 

To our knowledge, all previous global
inversion studies have used a flux representation defined on a
longitude-latitude grid \citep[or at continental scale,][etc.]{gurney2003,gurney2004}, with the grid centroids representing
flux values. The corresponding covariance
is defined based on distances between grid centroids, ignoring the
fact that the true underlying flux field is continuous. This
discrete representation is especially critical in cases where the grid centroids
represent spatial or temporal domains of different sizes,
such as the commonly used regular longitude-latitude grids.

In this article we use Gaussian Markov Random Fields (GMRFs) to define a spatially continuous flux model (Section~\ref{sec:Model}). The response from fluxes to observations is obtained by integrating fluxes on a continuous spatial domain (Section~\ref{sec:Traditional inverse modeling} and \ref{sec:observation_model}). Thus, aggregation errors related to the traditional assumption of constant flux within each grid cell are avoided. We use a separable spatio-temporal covariance function, where the temporal dependence in the fluxes is modelled by either an AR(1) or a seasonal AR(12) process (Section~\ref{sec:temporal_flux_model}). Both temporal dependence structures results in a sparse temporal precision matrix (inverse covariance matrix). The resulting flux model has a sparse spatio-temporal precision matrix on Kronecker form, which enables efficient computations. 

Our method is first demonstrated on pseudo data, with atmospheric
concentrations simulated from a known flux field (Section~\ref{sec:Simulation_study}).
Thereafter, the method is applied to real
atmospheric concentration data (Section~\ref{sec:Real data}) to reconstruct the global flux field between 1990 and 2001.  

\section{Atmospheric inverse modelling}
\label{sec:Traditional inverse modeling}
Atmospheric inverse modelling is based on the assumption that the trace gas concentrations can be determined by knowing all source and sink terms as well as the atmospheric transport of
the gas. If the
trace gas is inert, as $\text{CO}_2$, then an observation, $y_i=y(\bm{s}_i,t_i)$, of atmospheric trace
gas concentration  at location $\bm{s}_i\in S^2$ and time
$t_i\in \mathbb{R}$, can be seen as the contribution from the
continuous surface source flux, $f(\bm{s},t)$, integrated with respect to atmospheric transport, $J$, over past time $t \in (-\infty,\,t_i]$ and over
the Earth's surfaces $\bm{s} \in$ $S^2$:  
\begin{align}
\label{eq:observation_model0}
y_i 
 &= \int_{ -\infty}^{t_i} \int_{S^2} J(\bm{s}_i,t_i,\bm{s},t) f(\bm{s},t) d\bm{s} dt + \epsilon_{meas}(\bm{s}_i,t_i).
\end{align}
Here, $J(\bm{s}_i,t_i,\bm{s},t)$ quantifies the sensitivity of
$y(\bm{s}_i,t_i)$ with respect to the flux at spatial location
$\bm{s}$ and time $t < t_i$, and $\epsilon_{meas}(\bm{s}_i,t_i)$ is the
measurement error. The interpretation of $J$ is as the atmospheric
transport from flux location and time to the observation. After a
certain time, $K$, the surface source fluxes are assumed to be well mixed
in the atmosphere due to diffusion. Therefore, sensitivities to fluxes
at times $t< t_i-K$, are equal and given by a constant, $C$. The
constant sensitivity to older fluxes allows us to truncate the above
time integral at some initial time, $t_0$, and replace the response
from historical fluxes (fluxes at times $t<t_0$) with an initial atmospheric concentration $c_0$: 
\begin{align}
\label{eq:observation_model1}
y_i 
 &= c_0 + C \int_{t_0}^{t_i-K} \int_{S^2}  f(\bm{s},t) d\bm{s} dt + \int_{ t_i-K}^{t_i} \int_{S^2} J(\bm{s}_i,t_i,\bm{s},t) f(\bm{s},t) d\bm{s} dt + \epsilon_{meas}(\bm{s}_i,t_i).
\end{align}
In practise, the sensitivity $J{(\bm{s}_i,t_i,\bm{s},t)}$, is not
known, and needs to be approximated with a discrete representation
using an atmospheric transport model. The result is a transport model, $\bm{J}$,
that can be represented by a Jacobian matrix quantifying the sensitivity of observations to
discrete fluxes defined on a spatio-temporal grid; the transport
grid\footnote{$\bm{J}_{ij}=\frac{\partial y_i}{\partial
    f_j}$, where $f_j$ is an element of $\bm{f}$.}. In space, this is
commonly a regular longitude-latitude grid, whereas in time, the
resolution is typical monthly or daily. 

Traditionally, the integral
over the flux is discretized to the transport grid, with values at grid
centroids representing fluxes for the entire grid. The resulting discrete observation model can be presented in matrix form as 
\begin{align}
\bm{y} &= c_0 \cdot \bm{1} + \bm{J} \bm{f}+ \bm{\epsilon}, 
 \label{eq:observation_model_discrete} 
\end{align}
where $c_0 $ is the atmospheric \CO{}-concentration at time $t_0$, $\bm{1}$ is a unit vector, and $\bm{J}$ is the transport matrix, providing a linear link between surface fluxes, $\bm{f}$, and observed atmospheric concentrations, $\bm{y}$. A summary of notation can be found in Table~\ref{tab:symbols}.

Computational advantages arise by noting that the transport matrix can
be divided into two components, $\bm{J}=\begin{bmatrix} 
\bm{J}_c & \bm{J}_a \end{bmatrix}$,  based on the integral separation
in \eqref{eq:observation_model1}. Here, the "constant" part
of the transport matrix, $\bm{J}_c$, has a repeating pattern arising
from differences in grid cell volumes, whereas the "active" part of
the transport matrix, $\bm{J}_a$, has a band structure  that contains
information on concentration gradients. 

The observation errors, $\bm{\epsilon}$, incorporate both measurement
errors, $\bm{\epsilon}_{meas}$, and model errors,
$\bm{\epsilon}_{mod}$, where the latter are due to imperfections in the atmospheric transport model used to compute sensitivities. Since the
system in \eqref{eq:observation_model_discrete} is typically
under-determined, with more unknown flux elements than observations,
standard inversion techniques model the fluxes as spatio-temporal
Gaussian fields; introducing smoothing restrictions that give an
identifiable model.  
For grid-scale inversions, the fluxes are typically represented as
\begin{equation}
\label{eq:flux_representation_general}
f(\bm{s},t) = \mu(\bm{s},t) +  x(\bm{s},t),
\end{equation} 
with a expectation component, $\mu(\bm{s},t)$, and zero-mean spatio-temporal
Gaussian random field, $x(\bm{s},t)$. The expectation component (i.e.\ prior fluxes)  is commonly assumed known and given by a "best guess", based on estimates from
bottom-up modelling \citep[e.g.][]{kaminski1999b, rodenbeck2003}. 

As an alternative approach, \citet{michalak2004}, used a regression model for the expectation, given by
\begin{equation}
\label{eq:reg}
 \mu(\bm{s},t) =\sum_{j=1}^p B_j(\bm{s},t)\beta_j,
\end{equation}
where $\{B_j\}_{j=1}^p$ are known basis functions, and
$\{\beta_j\}_{j=1}^p$ are unknown regression coefficients. In this way, variables
believed to scale (linearly) with the \CO exchange, e.g.\ population and vegetation, could be included in the model.

The ability to constrain grid-scale inversions depends critically on
the assumed spatial and temporal correlations of the fluxes, as modelled
by the spatio-temporal random field, $x(\bm{s},t)$. Both
\citet{rodenbeck2003} and \citet{michalak2004}, among others,  
significantly reduce the flux uncertainties by modelling the spatial
dependence using exponential-like covariance functions.  A temporal
dependence with exponential covariance was introduced by \citet{rodenbeck2003}.

For the "large-scale" approaches, i.e., constant fluxes at continental
scale, the reduction in the number of unknown flux components often
results in identifiable models, even without the smoothing constraints of a spatial field. Excluding the field, $x(\bm{s},t)$,
from \eqref{eq:flux_representation_general}, these models reduce to linear regressions,
\eqref{eq:reg},  with indicator basis functions that specify the respective regions. 

\begin{table}[hbt]
\caption{Symbols and units (the following dimensions are true for modelling of a single flux field, and will change when both land- and ocean fluxes are modelled).}
\label{tab:symbols}
\begin{center}
\begin{tabular}{|c|c|c|c|}\hline
Symbol & Unit & Dimension & Attribute \\ \hline
$n_{s}$ &  --- & --- & $\#$Spatial elements in $\bm{f}$ \\
$n_{t}$ &  --- &  --- &  $\#$Temporal elements in $\bm{f}$ and $\bm{\omega}$ \\ 
$n_{\ell}$ &  ---& --- &  $\#$Triangular basis functions \\ 
$n_{\text{obs}}$ & --- & --- & Size of $\bm{y}$ \\
$p$ & --- & --- & Size of $\bm{\beta}$ \\
$n_{f} = n_t \cdot n_{s}$ & --- & ---  &  Size of $\bm{f}$ \\ 
$n_{\omega}= n_t \cdot n_{\ell}$ & --- & --- &  Size of $\bm{\omega}$ \\
$n_{z}=(n_t \cdot n_{\ell})+p$ & --- & --- &  Size of $\bm{z}$ \\
 \hline
$c_0$ & ppm & $1$ & Initial \CO concentration \\
$\bm{f}$ & kg/grid/year & $n_{f}$ & \CO flux on transport grid \\
$\bm{\omega}$ & kg/$\text{m}^2$/year & $n_{\omega}$ & Weights (GMRF) \\
$\bm{\beta}$ &  --- & $p$ & Regression coefficients \\
$\bm{z}$ &  kg/$\text{m}^2$/year & $n=n_{\omega}+p$ & Target (GMRF) \\
$\bm{J}$ & ppm/(kg/grid/year) & $n_{\text{obs}} \times n_{f}$ & Transport matrix \\
$\bm{H}$ & $\text{m}^2$/ $\text{grid}$ & $n_{f}\times n$ & Integration matrix \\
$\bm{A}$ & ppm/(kg/$\text{m}^2$/year) & $n_{\text{obs}} \times n$ & Observation matrix \\
 \hline
\end{tabular}
\end{center}
\end{table}

\section{Model}
\label{sec:Model}
The underlying continuous nature of the flux, and the steady increase of observations \citep[see e.g.][]{icos2019}, makes a continuous representation of the flux field attractive. 
This would minimize aggregation errors arising from the discrete grid
representation, and provide estimates of the flux covariance on a
continuous domain. Further a continuous model for fluxes allows the integration in \eqref{eq:observation_model1} to be performed at different spatial and temporal resolution for different flux components and regoins. This enables the combination of regional and global transport models, while maintaining a consistent definition of the flux covariance structure.

In this paper, the flux model \eqref{eq:flux_representation_general} is defined on a spatial continuous domain with discrete temporal resolution, using a Gaussian random field, $x(\bm{s},t)$. For completeness the model presented here allows for both a constant, $\mu_0(\bm{s},t)$, and a regression component, $\mu_\beta(\bm{s},t)$, in the expectation model, $\mu(\bm{s},t)$. However, in the application we only utilize constant prior fluxes. 

\subsection{Basis expansion}
A Gaussian random field can be specified through its covariance function, ${\text{C}(\bm{s},\bm{s}')}$. However, the spatial integration of fluxes in \eqref{eq:observation_model1} is problematic to compute for such a representation \citep[e.g.][suggests a solution based on Monte Carlo integration]{gelfand2010_gelfand}. Instead we model the spatial Gaussian field, in \eqref{eq:flux_representation_general}, at time $t$ using a basis expansion,
\begin{equation}
\label{eq:basis_expansion}
x(\bm{s},t)= \sum_{\ell=1}^{n_\ell} \phi_\ell(\bm{s})\omega_\ell(t),
\end{equation}
where $\phi_\ell(\bm{s})$ are basis functions defined on the sphere,
$\omega_\ell(t)$ are Gaussian weights, and $n_\ell$ is
the number of basis functions in the expansion. Stochastic fields
represented as in \eqref{eq:basis_expansion} have been
introduced before, typical for dimension reduction in applications on
big data sets with the aim of reducing computational
complexity. Examples include: basis expansion in the spectral domain using
spherical harmonics \citep{lang2015}, process convolutions
\citep{higdon2002}, and predictive process models
\citep{banerjee2008}. 
Basis functions with compact support have been used to obtain Markov fields that approximate stochastic fields with certain covariance functions \citep{lindgren2007, lindgren2011}. Regardless of the choice of basis functions in \eqref{eq:basis_expansion}, the spatial integration of the resulting random field, $x(\bm{s},t)$, over a spatial element, $\mathcal{D}$, is given by
\begin{eqnarray}
\int_{\bm{s}\in \mathcal{D}} x(\bm{s},t)ds  &=& \sum_{\ell=1}^{n_\ell} \left( \int_{\bm{s}\in \mathcal{D}} \phi_{\ell}(\bm{s})  d\bm{s} \right) \omega_{\ell}(t) = \bm{L} \bm{\omega}_{t},
\end{eqnarray}
where $\bm{\omega}_t =  \begin{bmatrix}
\omega_{1}(t) & \ldots  & \omega_{n_{\ell}}(t)
\end{bmatrix}^T$ are the weights at time $t$, and the linear
operator, $\bm{L}$, has elements given by the integration of the basis
functions, $L_{\ell}=\int_{\bm{s}\in \mathcal{D}} \phi_{\ell}(\bm{s})
d\bm{s} $. Assuming known basis functions which are independent of
model parameters, the elements of $\bm{L}$ can be precomputed \citep[see Appendix~\ref{App:Discretization} for technical details or e.g.][for examples of similar approaches]{SimpsILSR2016_online, MoragCMP2017_21}.

\subsection{Gaussian Markov Random Fields} 
For the application considered here, the spatial random field, $x(\bm{s},t)$ at time $t$ \eqref{eq:flux_representation_general}, will be modelled using the stochastic partial differential equation (SPDE) approach with piecewise linear basis functions defined on a Delaunay triangulation \citep{lindgren2011}. Gaussian Markov random fields on irregular grids were first introduced by \citet{lindgren2007} and extended to Markov representations of non-stationary fields by \citet{lindgren2011} and \citet{bolin2011}. The GMRFs are derived as weak solutions to the SPDE \citep{whittle1954},
\begin{equation}
(\kappa^2-\Delta)^{\alpha/2} \tau x(\bm{s},t)=\mathcal{W}(\bm{s}),
\label{SPDE}
\end{equation}
where $\Delta = \sum_{i=1}^d \frac{\delta^2}{\delta x_i^2}$ is the \textit{Laplacian} operator, $\mathcal{W}(\bm{s})$ is a Gaussian white noise process, $\kappa^2$ is a range parameter, and $\tau$ is a scaling parameter. The resulting stochastic fields have an approximate Mat\'ern covariance
\begin{equation*}
\label{Matern}
C(x(\bm{0},t),x(\bm{s},t)) \propto (\kappa \lVert \bm{s} \rVert)^{\nu}K_{\nu}(\kappa \lVert \bm{s} \rVert),
\end{equation*}
where $ \lVert \cdot \rVert$ represents the great circle distance on the sphere, $K_{\nu}$ is the modified Bessel function and $\nu>0$ and $\kappa>0$ are the regularity and range parameters, respectively. Further, $\alpha$ is related to $\nu$ as $\alpha = \nu + d/2$, where $d$ is the dimension of $\bm{s}$. The Mat\'ern covariance family is commonly used for geostatistical data due to its general form \citep{guttorp2006}, and it includes the exponential covariances ($\nu=1/2$) used in previous inversion studies.

Letting  $x(\bm{s},t)$ be a zero-mean Gaussian field expressed on the form \eqref{eq:basis_expansion}, solving the SPDE \eqref{SPDE}, with $\alpha=2$, results in the following distribution for the weights, $\omega_{\ell}(t)$,
\begin{align}
\label{eq:spatial_distribution}
\bm{\omega}_t 
& \in N(\bm{0},\bm{Q}_S^{-1}), 
& &\text{with} &
\bm{Q}_S &= \tau(\kappa^2\bm{C}+\bm{G})^T\bm{C}^{-1}(\kappa^2\bm{C}+\bm{G})\tau,
\end{align}
where $\bm{Q}_S$ is a sparse precision matrix \citep[see][for details]{lindgren2011} . 

\begin{figure}[t]
\begin{center}
   \includegraphics[width=\linewidth]{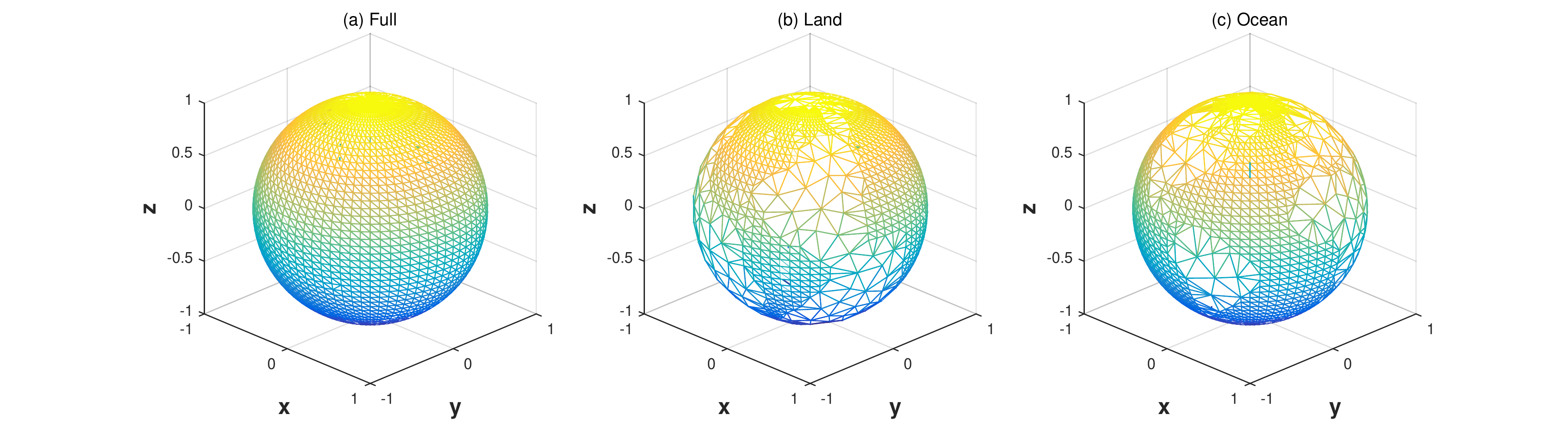}%
  \caption{Illustration of triangular basis functions on the sphere when modelling: (a) A single flux field, (b-c) a land and ocean flux field.}
  \label{fig:ALL_meshes}
  \end{center}
\end{figure}

Apart from a few constraints related to numerical stability the locations of the basis functions, and hence the resolution, can be specified freely \citep{BakkaRFRBIKSL2018_10}. When modelling land and ocean flux separately (see Appendix~\ref{App:Model extension}), we utilized this freedom by assigning a higher resolution for the domain of interest. The mesh for modelling a single flux field is displayed in Figure~\ref{fig:ALL_meshes} (a), whereas for separate flux fields for land and ocean, the meshes are illustrated in Figure~\ref{fig:ALL_meshes} (b--c).

\subsubsection{Spatio-temporal flux model}
\label{sec:temporal_flux_model}
The temporal dependence between fluxes is modelled using autoregressive processes. A spatio-temporal field with exponential covariance in time is obtained from a temporal AR(1) process, defined as
\begin{equation}
\label{eq:AR1}
\bm{\omega}_{t_m} = a\bm{\omega}_{t_{m-1}}+ \bm{\eta}_{t_m}
\end{equation}
with spatial dependent driving noise, $\bm{\eta}_{t_m}\in N\left(\bm{0}, \bm{Q}_S\right)$ \citep[Ch.~7]{blangiardo2015}. In addition, a seasonal dependence between fluxes is introduced by the following AR(12)-process, 
\begin{equation}
\label{eq:AR12}
\bm{\omega}_{t_m} = a\bm{\omega}_{t_{m-1}}+ b\bm{\omega}_{t_{m-12}} + \bm{\eta}_{t_m}
\end{equation}
where $b$ describes the temporal correlation between seasonal fluxes, e.g.\ January to January. The models described in \eqref{eq:AR1} and \eqref{eq:AR12}, together with \eqref{eq:spatial_distribution}, yield a latent field, $\bm{\omega} =  \begin{bmatrix}
\bm{\omega}_{t_1}^T & \ldots  &\bm{\omega}_{t_{n_{t}}}^T \end{bmatrix}^T$, with separable spatio-temporal dependence structure and sparse precision matrix \citep[Ch.~7]{blangiardo2015}. The distribution for $\bm{\omega}$ is given by
\begin{align}
\label{eq:dist_omega}
\bm{\omega} &\in N( \bm{0} ,\bm{Q}_\omega)
& &\text{with} &
\bm{Q}_\omega &= \bm{Q}_T(\bm{\zeta}) \otimes \bm{Q}_{S}(\bm{\theta})
\end{align} 
where $\otimes$ is the Kronecker product, $\bm{Q}_T$ is the temporal precision matrix from \eqref{eq:AR1} or \eqref{eq:AR12} (see supplementary material for details), $\bm{\zeta}=a$ or $\bm{\zeta} = \begin{bmatrix} a & b \end{bmatrix}$ details the temporal covariance parameters, and $\bm{\theta}= \begin{bmatrix} \tau & \kappa^2 \end{bmatrix}$ is a vector of spatial covariance parameters. 

\subsubsection{Marginal flux variance and range}
\label{sec:Marginal flux variance and range}
The covariance parameters are often easier to interpret when translated into marginal flux variance, $\sigma^2=V(x(\bm{s},t))$, spatial range, $\rho_S$, and temporal range, $\rho_T$. 
The spatial marginal variance, $\sigma^2_S$, is a function of the spatial covariance parameters $\kappa$ and $\tau$, given by
\begin{equation}
\sigma_S^2 = \frac{1}{\tau^2} \frac{\Gamma(\nu)}{\Gamma(\nu+d/2)(4\pi)^{d/2}\kappa^{2\nu}} 
\Biggr|_{\substack{\nu=1\\ d=2}}
= \frac{1}{4\pi \tau^2 \kappa^2}. 
\end{equation}
The temporal marginal covariance, $\sigma^2_T$, is obtained by solving the Yule-Walker equations \citep[see supplementary material and Ch.~8 in][]{Brock2009}. Following \citet{lindgren2011} the range is defined as the spatial/temporal distance at which the correlation is reduced to one tenth. The spatial range for a Mat\'ern covariance with $\nu=1$ is $\rho_S=\sqrt{8\nu}/\kappa=\sqrt{8}/\kappa$, the temporal range for an AR(1) process is $\rho_T = \log 0.1/\log a$, and for the AR(12) process the range is obtained numerically from the Yule-Walker equations. Note that the spatial distance is defined as the great circles distances divided by the Earth's radius, and the maximal distance between two locations is $\pi$ (or $180^\circ$). The temporal distance is defined in months.

\subsection{Flux and observation model}
\label{sec:observation_model}
The (numerical) integration of the different flux components \eqref{eq:observation_model1} to the transport grid can be described by a set of separate linear mappings, which depend on the spatial resolution of the different components; $\bm{L}_{0}$ for the constant expectation component $\mu_0(\bm{s},t)$, $\bm{L}_{\beta}$ for the regression component $\mu_\beta(\bm{s},t)$, and $\bm{L}_{\omega}$ for the spatially continuous random effect $x(\bm{s},t)$ (see Appendix~\ref{App:Discretization} for technical details). Resulting in
\begin{align}
\label{eq:flux_integration}
\bm{f} = \bm{L}_{0}\bm{\mu}_{0} +\underbrace{\begin{bmatrix}
 \bm{L}_{\omega} &  \bm{L}_{\beta}\bm{B} \end{bmatrix}}_\text{$\bm{H}$} \underbrace{\begin{bmatrix}
 \bm{\omega} \\  \bm{\beta} \end{bmatrix}}_\text{$\bm{z}$},
\end{align}
where $\bm{\mu}_{0}$ and $\bm{B}$ are discrete representations of the constant expectation and regression basis in \eqref{eq:reg}. Assuming a prior Gaussian field for 
$\bm{w}$, according to \eqref{eq:dist_omega}, and a non-informative Gaussian prior for the regression coefficients $\bm{\beta}$ the joint distribution of the unknown variables, $\bm{z}$, is
\begin{align}
\label{eq:target_model}
\bm{z}= \left[\begin{array}{c} 
\bm{\omega} \\ \bm{\beta}
\end{array}\right]  &\in N \left( \left[\begin{array}{c} 
\bm{0} \\ \bm{0} 
\end{array}\right],  {\underbrace{ \left[
\begin{array}{cc}
\bm{Q}_T(\bm{\zeta}) \otimes \bm{Q}_{S}(\bm{\theta}) & \bm{0}  \\ 
\bm{0} &  \bm{Q}_{\beta} 
\end{array} \right]}_{\bm{Q}_z}}^{-1} \right),
\end{align}
where $\bm{Q}_{\beta} = q_\beta \cdot \bm{I}$, and $q_\beta$ small (e.g.\ $q_\beta=10^{-4}$). The zero-valued off-diagonal elements in the precision matrix, $\bm{Q}_z$, arises from the assumption that $\bm{\omega}$ and $\bm{\beta}$ are a-prior independent. 

Combining the observation model \eqref{eq:observation_model_discrete} with the integration \eqref{eq:flux_integration} and the flux model defined in \eqref{eq:target_model}, we arrive at a conditional observation model 
\begin{align}
\label{eq:observation_model_Az}
\bm{y}|\bm{z} &\sim N \left( \bm{y}_{0} + \bm{A}\bm{z}, { \bm{Q}_{\epsilon}(\sigma^2_\epsilon)}^{-1}\right),
\end{align}
assuming Gaussian observational error, $\bm{\epsilon}$, with precision matrix $\bm{Q}_{\epsilon}(\sigma^2_\epsilon)$. Here, $\bm{y}_{0}  = c_{0} \cdot \bm{1} + \bm{J} \bm{L}_{0}\bm{\mu}_{0}$ is a deterministic expectation term with $c_{0}$ being the initial concentration, and $\bm{J} \bm{L}_{0}\bm{\mu}_{0}$ being the contribution from the fixed expectation. Moreover, $\bm{A}=\bm{J}\bm{H}$ with $\bm{H}$ from \eqref{eq:flux_integration}, is an observation matrix that maps the unknown variables, $\bm{z}$, to the observations, $\bm{y}$, by combining the integration in $\bm{H}$ and the atmospheric transport in $\bm{J}$. Similar to previous inversions, the observation errors, $\bm{\epsilon}$, are assumed to be independent, resulting in an error covariance matrix on the form 
\begin{equation}
\label{eq:observation_error}
\bm{Q}_{\epsilon}^{-1}(\sigma^2_\epsilon)= \sigma_\epsilon^{2}\bm{M}.
\end{equation} 
Here, $\bm{M}$ is a diagonal matrix with elements approximating the relative observational variance, and $\sigma_\epsilon^{2}$ is an unknown positive scaling constant.

\section{Estimation} \label{sec:Estimation}
This model, with a latent Gaussian field \eqref{eq:target_model}, and Gaussian observations \eqref{eq:observation_model_Az}, can be recognized as a standard model in the literature \citep[e.g.][p.~39]{rue2004}. The posterior distribution of $\bm{z}$ given observations and parameters is

\begin{equation}
\label{eq:posterior}
\bm{z}|\bm{y} \sim \mathcal{N}(\bm{\mu}_{\bm{z} | \bm{y}}(\bm{\Psi}),{\bm{Q}_{\bm{z}|\bm{y}}(\bm{\Psi})}^{-1}),
\end{equation}
where $\bm{\Psi}$ denotes all the (unknown) model parameters, and
the posterior expectation and precision are
\begin{align}
\label{eq:mu_rec}
\bm{\mu}_{{\bm{z}} | \bm{y}}(\bm{\Psi}) &= \bm{Q}_{\bm{z}|\bm{y}}^{-1}(\bm{\Psi}) \bm{A}^T\bm{Q}_{\epsilon}(\bm{\Psi})(\bm{y}-\bm{v_0}) \\
\label{eq:var_rec}
\bm{Q}_{\bm{z}|\bm{y}}(\bm{\Psi})&= \bm{Q}_{\bm{z}}(\bm{\Psi})+\bm{A}^T\bm{Q}_{\epsilon}(\bm{\Psi})\bm{A}.
\end{align}
Using \eqref{eq:flux_integration} posterior estimates and uncertainties for the fluxes are
\begin{align}
\label{eq:cond_fluxes_expectation}
\mathsf{E}(\bm{f} | \bm{y}) &= \bm{L}\bm{\mu}_{0} +\bm{H}\bm{\mu}_{{\bm{z}|y}}(\bm{\Psi}),\\
\mathsf{V}(\bm{f}|\bm{y}) &=\bm{H}\bm{Q}_{\bm{z}|\bm{y}}^{-1}(\bm{\Psi})\bm{H}^T. \label{eq:cond_fluxes_var}
\end{align}

The model parameters, $\bm{\Psi}$, are in general unknown, and estimates can be obtained by maximizing the likelihood \citep[see][for details]{rue2009}
\begin{equation}
\label{eq:INLA}
L_{\bm{\Psi}} = p(\bm{y}|\bm{\Psi}) = \frac{p(\bm{y}|\bm{z}, \bm{\Psi}) p(\bm{z}|\bm{\Psi})}{p(\bm{z}| \bm{y}, \bm{\Psi})}, \quad \forall \bm{z}.
\end{equation}
The expression in \eqref{eq:INLA} is valid for any $\bm{z}$, and a standard choice is to evaluate at $\bm{z}= \mu_{\bm{z} | \bm{y}}(\bm{\Psi})$ which reduces the likelihood to
\begin{equation}
  \label{eq:likelihood}
   L_{\bm{\Psi}} \propto 
   \left( \frac{\lvert \bm{Q}_{\bm{z}} \rvert \lvert \bm{Q}_{{\epsilon}} \rvert}{\lvert \bm{Q}_{\bm{z}|\bm{y}} \rvert} \right) ^{1/2} 
   \exp \left( -1/2 \left[ 
     \bm{\mu}_{\bm{z}|\bm{y}}^T \bm{Q}_{\bm{z}} \bm{\mu}_{\bm{z}|\bm{y}} +
     (\bm{y} - \bm{y}_{0}  - \bm{\mu}_{\bm{z}|\bm{y}})^T \bm{Q}_{{\epsilon}} 
     (\bm{y} - \bm{y}_{0}  - \bm{\mu}_{\bm{z}|\bm{y}}) \right] \right). 
\end{equation}
The posterior density \eqref{eq:posterior} is now obtained by maximizing the likelihood \eqref{eq:likelihood}, and using the estimated parameter in \eqref{eq:mu_rec} and \eqref{eq:var_rec}.
Note that $c_0$ in the mean component $\bm{y_0}$ is unknown, and is here estimated by averaging the first year of measurements after subtracting the response from the prior component, $\bm{J}\bm{L_0}\bm{\mu_0}$.

\subsection{Computational Issues}
For point observations of a latent field the observation matrix, $\bm{A}$, will be
sparse leading to a sparse posterior precision in \eqref{eq:var_rec}. Here the
inclusion of space-time integration in the observation model
\eqref{eq:observation_model0} leads to a dense observation matrix and dense
posterior precision matrix, due to the term
$\bm{A}^T\bm{Q}_{\epsilon}(\bm{\Psi})\bm{A}$. To avoid the $\mathcal{O}(n^3)$
computational cost associated with computing the inverses and determinantes of 
$\bm{Q}_{\bm{z}|\bm{y}}$ in \eqref{eq:mu_rec} and \eqref{eq:likelihood} we
simplify the expressions using matrix identities \citep{Harvi1997}.
Details of the simplifications and associated reductions in computational costs
can be found in the supplementary material; here we only summarise the results.

The cost of computing $\bm{Q}_{\bm{z}|\bm{y}}^{-1}$ in the posterior expectation
\eqref{eq:mu_rec} can be reduced by applying the Woodbury matrix identity. This
leads to a posterior variance expressed through the sparse precision matrices,
$\bm{Q}_{\epsilon}^{-1}$ and $\bm{Q}_z^{-1}$. The most expensive step when
computing the posterior expectation \eqref{eq:mu_rec} now reduces to solving the
equation system
\begin{align}
\label{eq:S}
  \bm{S} &= \bm{A}\bm{R}_{z}^{-1} = \bm{J}\bm{H}\bm{R}_{z}^{-1} =
  \begin{bmatrix} \bm{J}_c & \bm{J}_a \end{bmatrix} \bm{H}\bm{R}_{z}^{-1},
 & &\text{where} &
 \bm{Q}_z &= \bm{R_z}^T\bm{R_z}
\end{align}
and $\bm{R}_{z}$ is the Cholesky factorization of
$\bm{Q}_z$. Using the block and Kronecker structure of $\bm{Q}_z$ and the division of $\bm{J}$ into constant and active parts the cost of computing $\bm{S}$ is
$\mathcal{O} (n_{\text{obs}} n)$ since $\bm{R}_z$ is almost as sparse as $\bm{Q}_z$ \citep[After reordering $\bm{R}_z$ will have $\mathcal{O} (n_t n_\ell \log n_\ell)$ non-zero elements, see supplementary material and][p.~51.]{rue2009}.

The determinants in the likelihood \eqref{eq:likelihood} can be simplified
\citep[Thm.~8.1]{Harvi1997} to
\begin{equation}
\label{eq:det}
\frac{\lvert \bm{Q}_{{\bm{z}}} \rvert \lvert \bm{Q}_{\epsilon} \rvert}{
 \lvert \bm{Q}_{\bm{z}|\bm{y}} \rvert} = 
 \frac{1}{\lvert \bm{Q}_{\epsilon}^{-1} +  \bm{S}\bm{S}^T \rvert},
\end{equation}
where the most expensive calculations is due to the
$\mathcal{O} (n_{\text{obs}}^2 n)$ cost of the $\bm{S}\bm{S}^T$-product. Finally we introduce $\bm{L}$ as the following Choleskey factorization
\begin{align*}
\bm{L}^T \bm{L} &= (\bm{Q}^{-1}_{\epsilon}+\bm{S}\bm{S}^T).
\end{align*} 

The resulting simplified posterior expectation and negative log likelihood are:
\begin{align}
  \label{eq:new_posterior_expectation}
  \bm{\mu}_{{z|y}} &=\bm{R}_{z}^{-1}( \bm{M}- \bm{S}^T \bm{L}^{-1} \bm{L}^{-T} \bm{S} \bm{M})
  \\
\label{eq:new_likelihood}
 - \log L_{\bm{\Psi}}  & \propto  \log \lvert \bm{L} \rvert +  
\frac{1}{2} \left( \bm{\mu}_{{z|y}}^T \bm{Q}_{\bm{z}} \bm{\mu}_{{z|y}} + 
(\bm{y} - \bm{y}_{0}  - \bm{\mu}_{\bm{z}|\bm{y}})^T \bm{Q}_{{\epsilon}}
(\bm{y} - \bm{y}_{0}  - \bm{\mu}_{\bm{z}|\bm{y}})\right),
\end{align}
where ${\bm{M}=\bm{S}^T (\bm{Q}_{\epsilon} (\bm{y}-\bm{y}_{0}))}$.

\subsection{Computation of posterior variances}
Inverting the posterior precision matrix in \eqref{eq:var_rec} to obtain marginal posterior variances for the conditional fluxes \eqref{eq:cond_fluxes_var} is prohibitively expensive. Instead we use a sampling-based approach \citep{bekas2007}, for which the diagonal of a matrix $\bm{B}$, is approximated with the  unbiased estimator:
\begin{equation}
\label{eq:Bekas}
\widehat{\text{diag}(\bm{B})} = \bm{U} \odot (\bm{B} \bm{U}).
\end{equation}
Here, $\odot$ is the element-wise product, and $\bm{U}$ is a random column vector with elements sampled independently from the Rademacher random numbers; $\mathsf{P}(\bm{U}= \pm 1) = 0.5$. Note that the estimate in \eqref{eq:Bekas} does not require us to form the complete matrix $\bm{B}$, only the ability to compute $\bm{B}$ multiplied by a vector.
Combining \eqref{eq:Bekas} and \eqref{eq:cond_fluxes_var} we obtain an estimate of the diagonal elements of the conditional variance, given by: 
\begin{equation}
\label{eq:Bekas_f}
\widehat{\text{diag}(\mathsf{V}(\bm{f}|\bm{y}))} = \bm{U} \odot (\bm{H}\bm{Q}_{{z|y}}^{-1}\bm{H}^T \bm{U}).
\end{equation}
The uncertainty in the estimator is reduced by averaging over $10\,000$ independent vectors, $\bm{U}$. Again, using the Woodbury matrix identity, we obtained a computationally faster expression,
\begin{equation}
\label{eq:Bekas_f_simplified}
\widehat{\text{diag}(\mathsf{V}(\bm{f}|\bm{y}))} = \bm{U} \odot  \left[\bm{H}(\bm{R_z}^{-1}(\bm{I}-\bm{S}^T\bm{L}^{-1}\bm{L}^{-T}\bm{S}) \bm{R_z}^{-T}\bm{H}^T \bm{U}\right].
\end{equation}
Here $\bm{I}$ is the identity matrix, and the matrix products can be evaluated in the computationally most beneficial order.

\section{Application}
In this section, we apply our method to a real global atmospheric inversion problem. 
To investigate the full potential of our inversion system, our first analysis is based on simulated \CO concentrations (at time and locations where we have real data) obtained by running the transport model forward assuming known fluxes (Section~\ref{sec:Simulation_study}). 
Secondly, using the same inversion system, the posterior fluxes \eqref{eq:flux_integration} are estimated using real measurements of \CO concentrations (Section~\ref{sec:Real data}). 

In both studies, monthly flux fields of \CO are estimated for the
period 1990 to 2001, using observations of \CO concentration
from 1992 to 2001. Thus, each of the observations will have a model
response to at least 25 months of fluxes. The lack of observations in
1990-1991 make the fluxes in this period badly constrained. Therefore,
(estimated) posterior fluxes will be analysed for the period 1992 to 2000.

\subsection{Data}
\label{sec:Data}
\subsubsection{Observations} \label{subsec:Observation}
The atmospheric CO$_2$ data are generated according to the procedure
described by \citep{rodenbeck2005} for the Jena Carboscope. It is based
on samples collected and analysed by several institutions. Monthly
mean values are calculated as the average of the 
measurements taken within the selected month. 

\begin{figure}[htb]
  \centerline{ \includegraphics[width=.8\linewidth]{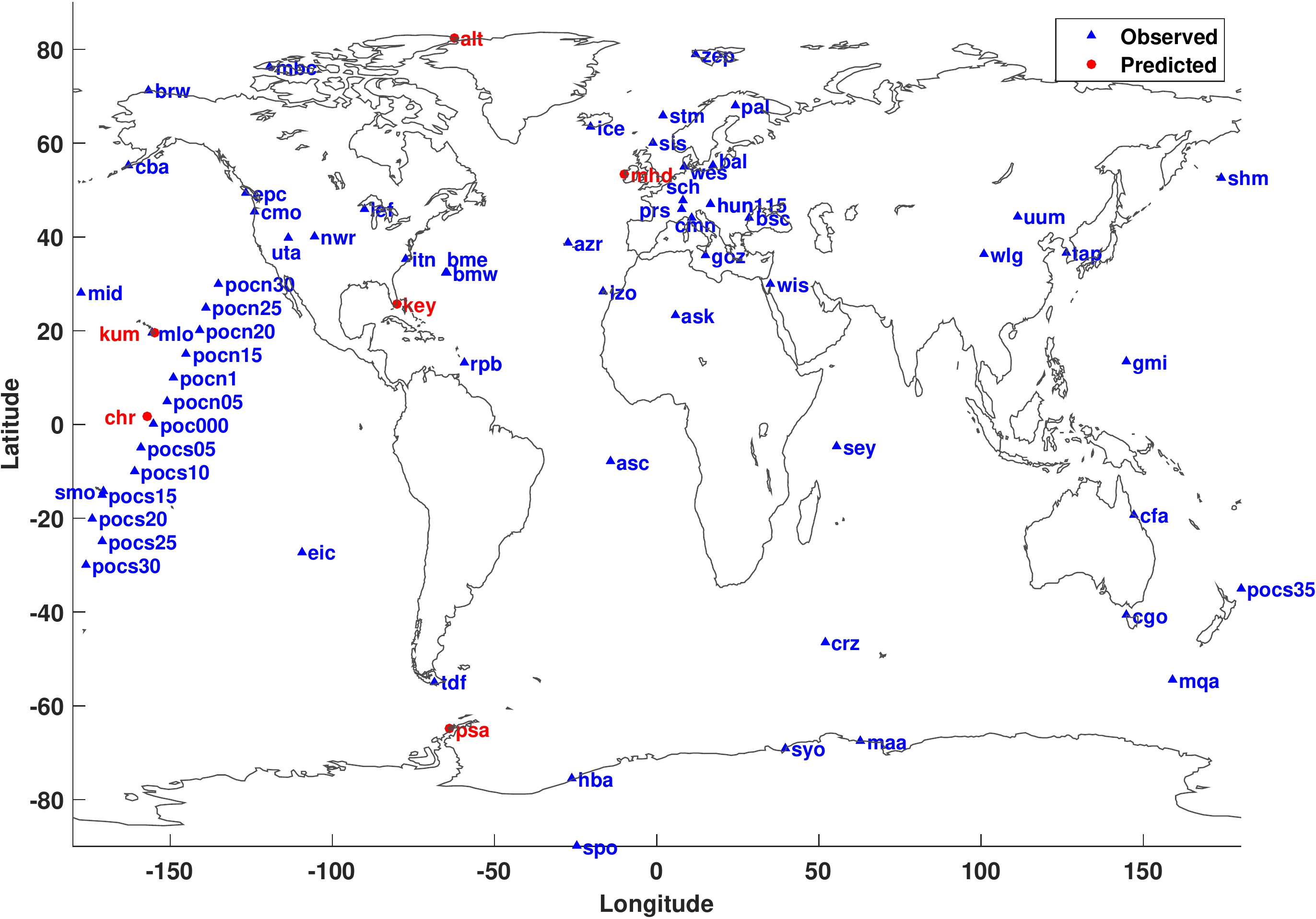} }%
  \caption{Network of global measurement stations. The observations locations marked with blue triangles are used to estimate fluxes, whereas observation locations indicated with red circles is used for validating the different model performances.}
  \label{fig:observation_location}
\end{figure}

The spatial network of \CO monitoring sites is illustrated in Figure~\ref{fig:observation_location}. Here, locations indicated with blue triangles are used for estimating the fluxes, while locations marked with red circles are used for validation. The number of available observations at each station varies in time and is shown in Appendix~\ref{App:Additional material}, Figure~\ref{fig:observation_time}. 

\subsubsection{Measurement error}
\label{subsec:Measurement error}
The measurement error is expressed on the form \eqref{eq:observation_error}, and in the application on real data, the diagonal elements of $\bm{M}$ is defined by
\begin{equation}
\label{eq:M}
M_{ii} = \sigma_{mod,i}^2 + \sigma_{meas,i}^2.
\end{equation}
As in \citet{rodenbeck2005}, the model error, $\sigma_{mod}$, is classified based on the location type, see Table~\ref{tab:classification}. In our study, the measurement error, $\sigma_{meas}$, is set to $0.3$ ppm independently of the number of raw concentration averaged to obtain $y_i$. This simplification will have a small effect on the elements of $\bm{M}$, since the model error is much larger than the measurement error. The class specification for the 70 different measurement locations are given in the supplementary material. 
\begin{table}[htb]
\caption{Class-separated model error that depends on the geographical location as proposed by \citet{rodenbeck2005}.}
\label{tab:classification}
\begin{center}
\begin{tabular}{llc}
\hline
\textbf{Class} & \textbf{Description} & \textbf{$\sigma_{mod}$ (ppm)} \\ \hline
C & Continental & 3.0 \\
M & Mountain & 1.5 \\
R & Remote & 1.0 \\
S & Shore & 1.5
\end{tabular}
\end{center}
\end{table}
\subsubsection{Transport model}
The meteorological transport of fluxes is here approximated by the
atmospheric tracer transport model TM3 \citep{heimann2003}, with a
latitude-longitude resolution of approximately $3.8^{\circ}\times
5^{\circ}$, yielding $48 \times 72$ rectangular grid cells, and 19
vertical levels. As mentioned before,
Jacobian representations of the atmospheric transport model
TM3 are used in this study. The monthly Jacobians were calculated by
sampling the transport model according to the timing of the actual
measurements. The time resolution is one month, and information on gradient concentration range over the last $C=36$ months \eqref{eq:observation_model1}. The sensitivity to fluxes further back in time takes a constant value.  As there is no terrestrial productivity on Antarctic and Greenland, the sensitivities to fluxes at these locations are set to zero. 
\subsubsection{Triangulation}

The resolution of the triangulation is similar to the resolution of the TM3 transport grid. The triangulation is obtained by letting the node locations of the basis functions, $s_\ell$, be centred in the grid cells of the TM3 model (apart from the grid cells on the poles, where the nodes are located more sparse). This prevents the resolution specified by the triangular nodes to be lower than what can be resolved by the transport matrix. The mesh for modelling a single flux field is shown in Figure~\ref{fig:ALL_meshes} (a). When modelling separate land and ocean fields, the flux resolution is higher for the spatial domain of interest, as illustrated in Figure~\ref{fig:ALL_meshes} (b--c). Any node included in the mesh for a single flux can be found in either the land or ocean mesh. Moreover, node-points on the boundaries, i.e., node-points centred in TM3 grid cells containing both land and ocean, are included in both the land and ocean triangulations.

\subsubsection{Prior fluxes}
\label{sec:Prior fluxes}
Prior fluxes consist of three components: net
land-atmosphere exchange due to vegetation (Net Ecosystem Exchange, NEE), net
ocean-atmosphere exchange (ocean flux), and fossil fuel emissions. All
components are resolved on the transport grid (with a spatial
resolution of $\approx4^{\circ}\times 5^{\circ}$). The NEE fluxes consist of monthly NEE
simulated from the BETHY dynamic vegetation model
\citep{knorr2000}, whereas the
ocean fluxes are temporally flat, with spatial ocean fluxes taken from
\citet{takahashi2002}. Fossil fuel emission fields are resolved on a monthly time-scale with fossil fields obtained
based on linear combinations of the emission fields evaluated at year
1990 \citep{andres1996} and 1995 \citep{brenkert1998}.

\subsection{Models}
In total six models (see Table~\ref{tab:models}) are introduced for modelling the latent field; S0, S1, S12, B0, B1 and B12. The letter S denotes models that include a single flux field, and thus do not separate ocean- and land-flux dynamics. The letter B denotes models that include both land and ocean flux fields (see Appendix \ref{App:Model extension}). The model number specifies the order of the autoregressive model used for modelling the time dependence. For example, model S0 models a single flux field assuming temporal independence. 
\begin{table}[htb]
\caption{Description of the six models used.}
\label{tab:models}
\begin{center}
\begin{tabular}{lcc}
\hline
\multicolumn{1}{l}{\textbf{Model}} & \multicolumn{1}{c}{\textbf{Field}} & \multicolumn{1}{l}{\textbf{AR-order}} \\ \hline
S0 & Single & 0 \\
S1 & Single & 1 \\
S12 & Single & 12 \\
B0 & Separate Land and Ocean & 0 \\
B1 & Separate Land and Ocean  & 1 \\
B12 & Separate Land and Ocean  & 12 \\
\end{tabular}
\end{center}
\end{table}

\subsubsection{Simulation study}
\label{sec:Simulation_study}
In the simulation study we only consider a model with zero-mean and a single spatio-temporal random field with seasonal dependence, i.e.\ model $\text{S12}$. For the simulated observations, two main cases are considered.

In the first case, referred to as \textit{bottom-up fluxes}, pseudo-observation are based on the prior NEE and ocean fluxes (see Section~\ref{sec:Prior fluxes}); excluding the fossil fuel component. The aim of using \CO observations simulated from the bottom-up fluxes is to investigate the quality of the reconstructions based on known fluxes that resemble real-world \CO fluxes, and might not follow the Gaussian (and other) assumptions of the model. 

In the second case, referred to as the \textit{Gaussian fluxes}, the known fluxes are obtained by simulating the latent field from the stochastic model S12 using  parameters estimated from the prior NEE and ocean fluxes\footnote{By solving $\bm{f}=\bm{L}_{\omega}\bm{\omega}$, with $\bm{f}$ being the NEE and ocean fluxes; the parameters were obtained based on a standard maximum likelihood estimation: $[\hat{\bm{\zeta}}, \hat{\bm{\theta}}] =  \underset{\bm{\zeta},\bm{\theta}} {\argmax} \  p(\bm{w}|\bm{\zeta},\bm{\theta})$.}. The aims of using simulated Gaussian fluxes are to evaluate how well model parameters are constrained by the observations and to compare reconstructions obtained with estimated versus true parameters.

In both cases, pseudo-observations are simulated according to \eqref{eq:observation_model_discrete}, with observation error defined on the form \eqref{eq:observation_error}. Two different noise levels are added; "low" and "high", resulting in four sets of simulated data. The low observation noise has a standard error of $0.1$ ppm ($M_{ii}=0.01$, and $\sigma_\epsilon=1$), whereas the the high observational noise has standard deviations in the range 1-3 ppm (with $M$ defined in \eqref{eq:M}, and $\sigma_\epsilon=1$). The "high" observational noise approximates the variance of the real observations, while the "low" noise should highlight the reconstruction limits of the transport model.

\subsubsection{Real data}
\label{sec:Real data}
The real fluxes are modelled using a deterministic expectation, $\mu_0(\bm{s},t)$, and a spatio-temporal field, $x(\bm{s},t)$. The expectation or prior flux, $\mu_0$, is constructed by adding up contributions from: 1) a seasonal NEE component obtained by averaging each calender month during 1982--1990; 2) ocean fluxes from \citet{takahashi2002} which are already averaged in time; and 3) fossil fuel emissions. For the spatio-temporal field all six models in Table~\ref{tab:models} are considered.

\section{Results}
\label{sec:Results}

\subsection{Simulation study}
\begin{figure}[t]
\centering
\includegraphics[width=0.99\linewidth]{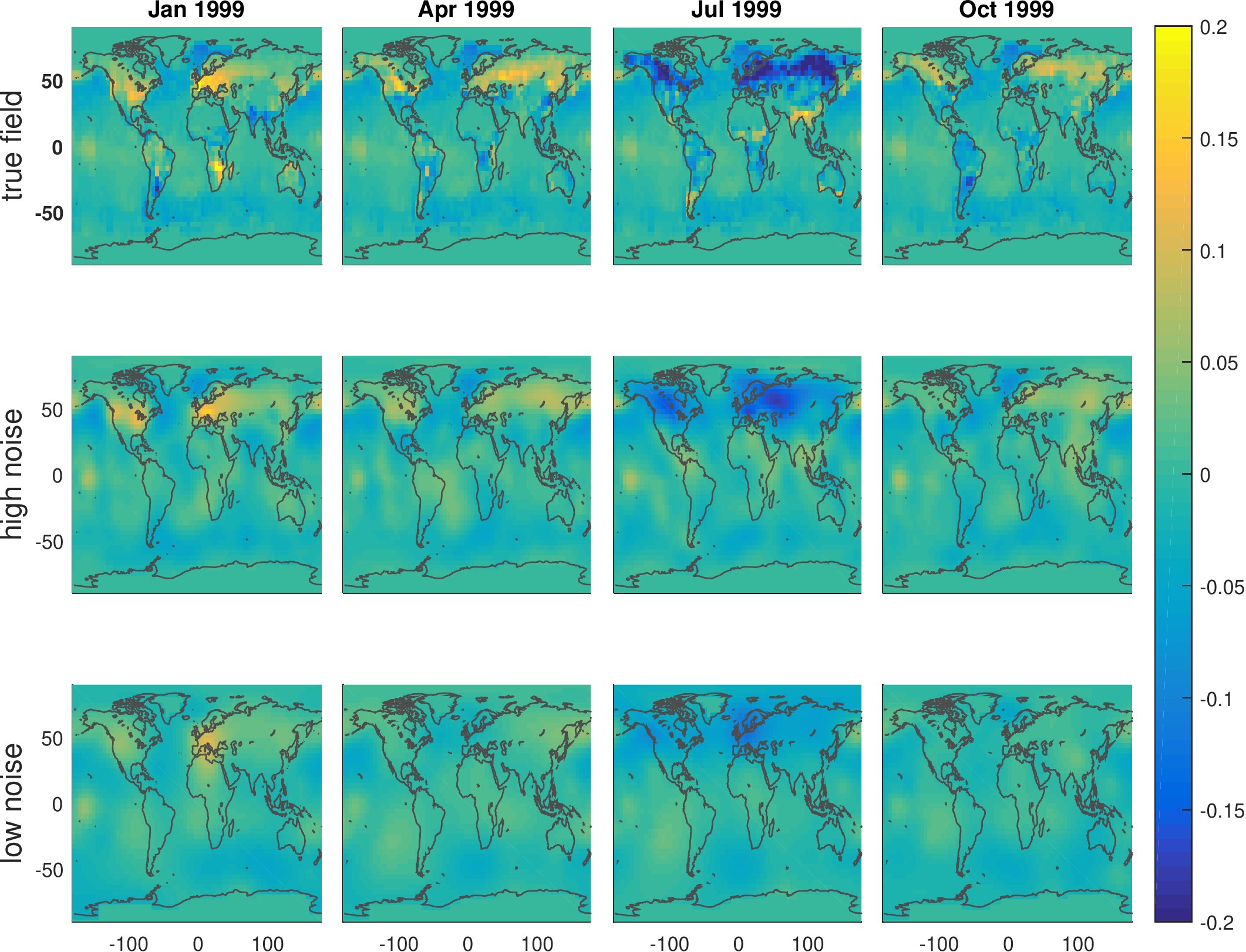} %
  \caption{Results for the simulation study using observations simulated from the bottom-up fluxes. The first row shows the actual \CO bottom-up fluxes [$\text{kgC}/(\text{year}\cdot \text{m}^2)$] for January, April, July, and October of 1999. The second and third row show the resulting posterior fluxes estimated using the S12-model and observations simulated with low and high noise, respectively.}
   \label{fig:rec_pseudoData}
\end{figure}

For the simulation study, parameter estimates are provided in Table~\ref{tab:true_parameters}; root mean square error (RMSE) between the true and  posterior flux fields, as well as between predicted and observed \CO concentrations at the validation sites are given in Table~\ref{tab:RMSE_sim}; and reconstructed fluxes for the bottom up case are given in Figure~\ref{fig:rec_pseudoData} (for January, April, July and October of 1999).

For the Gaussian fluxes the estimated parameters are close to those used to generate the data (Tbl.~\ref{tab:true_parameters}); especially for low observational noise. For high observational noise, the estimated range remains close to the true, while the scaling parameter, $\tau$, and the temporal parameters, $a$ and $b$, are slightly shifted. In the case of bottom-up fluxes, the deviations between parameters estimated directly from the fluxes ("true" parameters) and those estimated from \CO observations are large; we specifically note the shift towards longer spatial range (smaller $\kappa$ values) and stronger seasonal dependence (large $b$) when basing estimates on \CO observations.
These discrepancies might be due to deviations from our Gaussian assumptions. Despite this, the reconstruction errors are of similar magnitude as for the simulated Gaussian fields (Tbl.~\ref{tab:RMSE_sim}).

Comparing RMSE-values for both \CO observations at validation sites and the complete flux fields, the errors are comparable between the Gaussian and bottom up data. Recall that the Gaussian fields are simulated using parameters estimated from the bottom up fields and should have comparable variances. Errors increase slightly when using high noise, and for all cases estimated parameters perform as good as or better than known parameters.
 
The inverse system performs quite well for the low observational noise. Some spatial flux information is lost across the Southern hemisphere and Siberia (Fig.~\ref{fig:rec_pseudoData}); i.e.\ across regions with few measurement locations (Fig.~\ref{fig:observation_location}). Using the pseudo-data with the higher noise level, a lot of spatial structure is smoothed out in the reconstructed fluxes, and strong signals in the bottom-up fluxes are only captured when located close to observation sites. However, the posterior fluxes still resolve the large scale patterns of the true fluxes, especially for the Northern hemisphere.

\begin{table}[htb]
\caption{Estimated parameter in the simulation study. "low/high" refers to estimated parameters based on \CO concentrations observed with low and high noise, respectively. "True" refers to true parameters, i.e.\ those estimated from the bottom up flux fields and then used to simulate the Gaussian fluxes. Spatial and temporal range, and standard deviation of the latent field corresponding to the estimated parameters are provided in the lower part of the table.}
\label{tab:true_parameters} 
\begin{center}
\begin{tabular}{lS|SS|SS}
\cline{2-6}
& \multicolumn{1}{c|}{\textbf{True}} & 
\multicolumn{2}{c|}{\textbf{Bottom up fluxes}} &
\multicolumn{2}{c}{\textbf{Gaussian fluxes}} \\ 
\textbf{Parameters} & & Low & High & Low & High \\ \hline
$\sigma_\epsilon$ &    1.00 &    1.28 &   0.995 &    1.00 &  1.05 \\
$\tau$            &   0.447 &    18.7 &    5.30 &   0.402 & 0.303 \\
$\kappa$          &    23.9 &    3.53 &    5.18 &    22.9 &  23.3 \\
$a$               &   0.555 &  0.0108 &   0.103 &   0.494 & 0.276 \\
$b$               &   0.326 &   0.989 &   0.889 &   0.272 & 0.370 \\ \hline
$\rho_S$ -- spatial r.  &  0.118 &   0.801 &  0.546 &  0.124 & 0.121 \\
$\rho_T$ -- temporal r. &   65.0 &{$>10^4$}&   2060 &   27.6 & 25.6 \\
$\sigma$ -- std.\ dev.  & 0.0391 &  0.0734 & 0.0372 & 0.0387 & 0.0458 
\end{tabular}
\end{center}
\end{table}

\begin{table}[htb]
\caption{RMSE calculated for flux fields and observations (at validation sites) using either true (i.e.\ estimated directly from the bottom up flux fields or used to simulate the Gaussian fluxes) or estimated (from simulated \CO observations) model parameters. The results in the first two columns are obtained by simulating observations from bottom-up fluxes, while the results in the last two columns are obtained by simulating Gaussian fluxes. The notation "low" and "high" refers to low and high observational noise, respectively.}
\label{tab:RMSE_sim}
\begin{center}
\begin{tabular}{ll|SS|SS}
\cline{2-6}
& & \multicolumn{2}{c|}{\textbf{Bottom-up fluxes}} & 
\multicolumn{2}{c}{\textbf{Gaussian fluxes}} \\
& Parameters & Low & High & Low & High \\ \hline
\multirow{2}{*}{\textbf{\begin{tabular}[c]{@{}l@{}}RMSE fluxes \\
($\text{kgC}/(\text{year}\cdot \text{m}^2)$)\end{tabular}}} & True &
0.0272 & 0.0319 & 0.0401 & 0.0410
\\
& Estimated & 0.0245 & 0.0294 & 0.0400 & 0.0410
\\ \hline
\multirow{2}{*}{\textbf{\begin{tabular}[c]{@{}l@{}}RMSE val. data \\ 
(ppm)\end{tabular}}} & True & 
0.245 & 0.508 & 0.518 & 0.579
\\
& Estimated & 0.143 & 0.406 & 0.511 & 0.568               
\end{tabular}
\end{center}
\end{table}

\subsection{Real data}
Model performance for real data was evaluated by computing information criterias, AIC \citep{AIC_Akaike} and BIC\footnote{$\text{AIC} = 2k-2\log L_{\bm{\Psi}}$, and $\text{BIC} =  k \log(n_{\text{obs}})-2\log L_{\bm{\Psi}}$, where $k$ is the number of parameters in $\bm{\Psi}$ and $L_{\bm{\Psi}}$ is the maximum value of the likelihood.} \citep{schwarz1978}, as well as RMSE for the validation data (Tbl.~\ref{tab:information_criterias}). Estimated model parameters for all six models are listed in Table~\ref{tab:model_parameters}.

\begin{table}[htb]
\caption{Information criterias (AIC \& BIC) and root mean square error (RMSE) calculated for the prior fluxes, and posterior fluxes estimated using the six different models; lower values are better.}
\label{tab:information_criterias}
\begin{center}
\begin{tabular}{c|ccS}
\hline
\textbf{Model} & \textbf{AIC} & \textbf{BIC} & {\textbf{RMSE} (ppm)} \\ \hline
$\mu_0$ & & & 2.79 \\
S0 & 23\,819 & 23\,839 & 1.40 \\
S1 & 22\,587 & 22\,614 & 1.51 \\
S12 & 19\,651 & 19\,684 & 1.62 \\
B0 & 20\,896 & 20\,930 & 1.33 \\
B1 & 19\,590 & 19\,636 & 1.35 \\
B12& 17\,995 & 18\,054 & 1.31
\end{tabular}
\end{center}
\end{table}

\begin{table}[htb]
\caption{The estimated model parameters for all six models are provided in the top part of the table. The bottom part of the table gives the corresponding spatial and temporal ranges, and marginal standard deviations.}
\label{tab:model_parameters}
\begin{center}
\begin{tabular}{l|SSS|SSS}
\hline
\textbf{Parameter} & {\textbf{S0}} & {\textbf{S1}}  & {\textbf{S12}}  & {\textbf{B0}} & {\textbf{B1}}  & {\textbf{B12}}\\ \hline
$\sigma_{\epsilon}$ & 1.05  &  0.921  &  0.904 & 0.824  &  0.860  &  0.897\\
$\tau_{la}$ & 1.31 &   0.265 & 0.285 &  0.00705 & 0.0137 & 0.0239\\
$\kappa_{la}$ & 3.72 & 11.6 & 15.5 & 38.6 & 30.5 & 36.7 \\
$a_{la}$ &  & 0.782 & 0.269 &  & 0.618  &  0.149\\
$b_{la}$&   &   &  0.723 & & & 0.832\\
$\tau_{oc}$ &  &   &  & 0.422 & 0.716 & 1.19\\
$\kappa_{oc}$ &  &   &  & 10.8 & 24.2 & 13.3\\
$a_{oc}$ &  &   &  &  & 0.987 & 0.121 \\
$b_{oc}$&   &   &  &   &  &  0.870\\ \hline
$\rho_{S,la}$ -- spatial r.  & 0.760  & 0.244  & 0.183 & 0.0733 &  0.0928 &  0.0771\\
$\rho_{T,la}$ -- temporal r. &  & 9.36 & 2140 &    & 4.78 &  767 \\ 
$\sigma_{la}$ -- std.\ dev.  & 0.0579  &  0.147  &  0.206  &  1.04  &  0.859  &  0.858 \\
$\rho_{S,oc}$ -- spatial r.  &  &  &  &  0.262 &  0.117 & 0.213 \\
$\rho_{T,oc}$ -- temporal r. &  &  &  &   &  176   &  1810 \\
$\sigma_{oc}$ -- std.\ dev.  &  &  &  &  0.0619  &  0.101   & 0.0606 
\end{tabular}
\end{center}
\end{table}

Including separate models for land and ocean improves performance across all three metrics (AIC, BIC, and RMSE), regardless of the temporal dependence. This is to be expected given the very different dynamics of land and ocean fields; illustrated by the differences in estimated marginal standard deviations, $\sigma_{la}$ and $\sigma_{oc}$, and ranges, $\rho_{la}$ and $\rho_{oc}$. Increasing the temporal structure in the model substantially decreases AIC and BIC for both the single and separate field cases. However, the RMSE values do not follow such a simple pattern. For the models with both land and ocean fields including a seasonal, i.e.\ AR(12), dependence gives the best result for the validation data, while the models with a single field perform best with no temporal dependence (the RMSE values at each observation and validation site can be found in the supplementary material). Overall the B12 model, land and ocean fields with seasonal dependence, performs best across all metrics.

\begin{figure}[htb]
\begin{center}
   \includegraphics[width=0.99\linewidth]{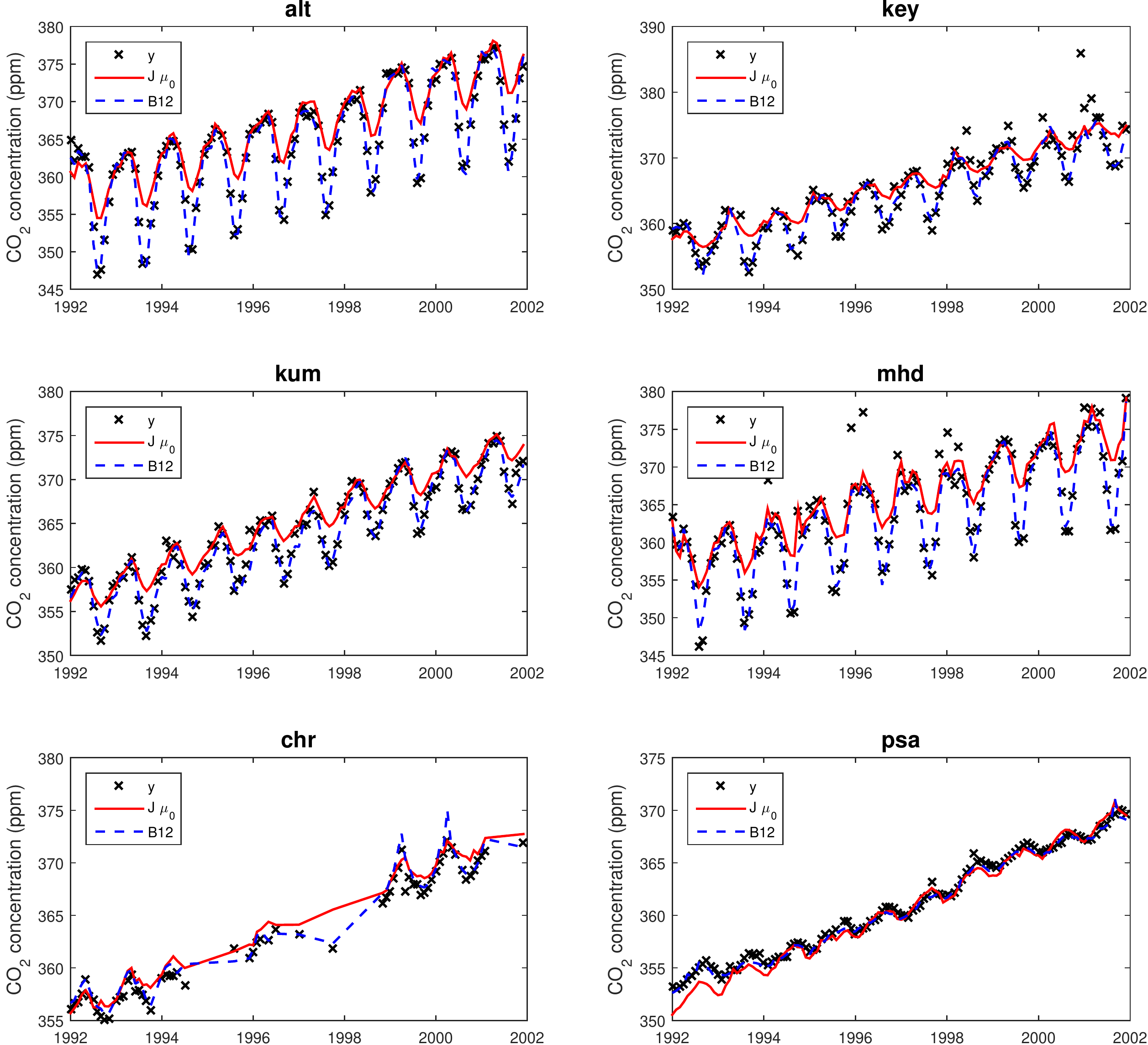}%
  \caption{Observed atmospheric \CO concentrations (black crosses) at the six validation sites (see Fig.~\protect{\ref{fig:observation_location}}), illustrated together with predicted concentrations based on the prior fluxes (red solid line) and posterior fluxes (blue dashed line) obtained from the best model: B12.}
   \label{fig:val_trueData}
  \end{center}
\end{figure}

The improvement from introducing a seasonal dependence is likely due to the strong seasonal trend present in the observational data. In Figure~\ref{fig:val_trueData}, the predicted concentrations at the validation sites based on the B12-model are illustrated together with the observations and the concentrations due to prior fluxes, $\bm{y}_{0}=\bm{c}_{0}+\bm{J}\bm{\mu}_{0}$ in \eqref{eq:observation_model_Az}. The prior fluxes fail to capture the reduction in \CO concentrations during summer (June, July, and August), resulting in a strong seasonal component that needs to be modelled by the latent field. We note the generally excellent agreement of the posterior concentrations estimated using the B12 model with the observations, except for some outliers at mainly Mace Head, Ireland (mhd) and Key Biscayne, Florida (key). 

\begin{figure}[htb]
\centering
   \includegraphics[width=0.95\linewidth]{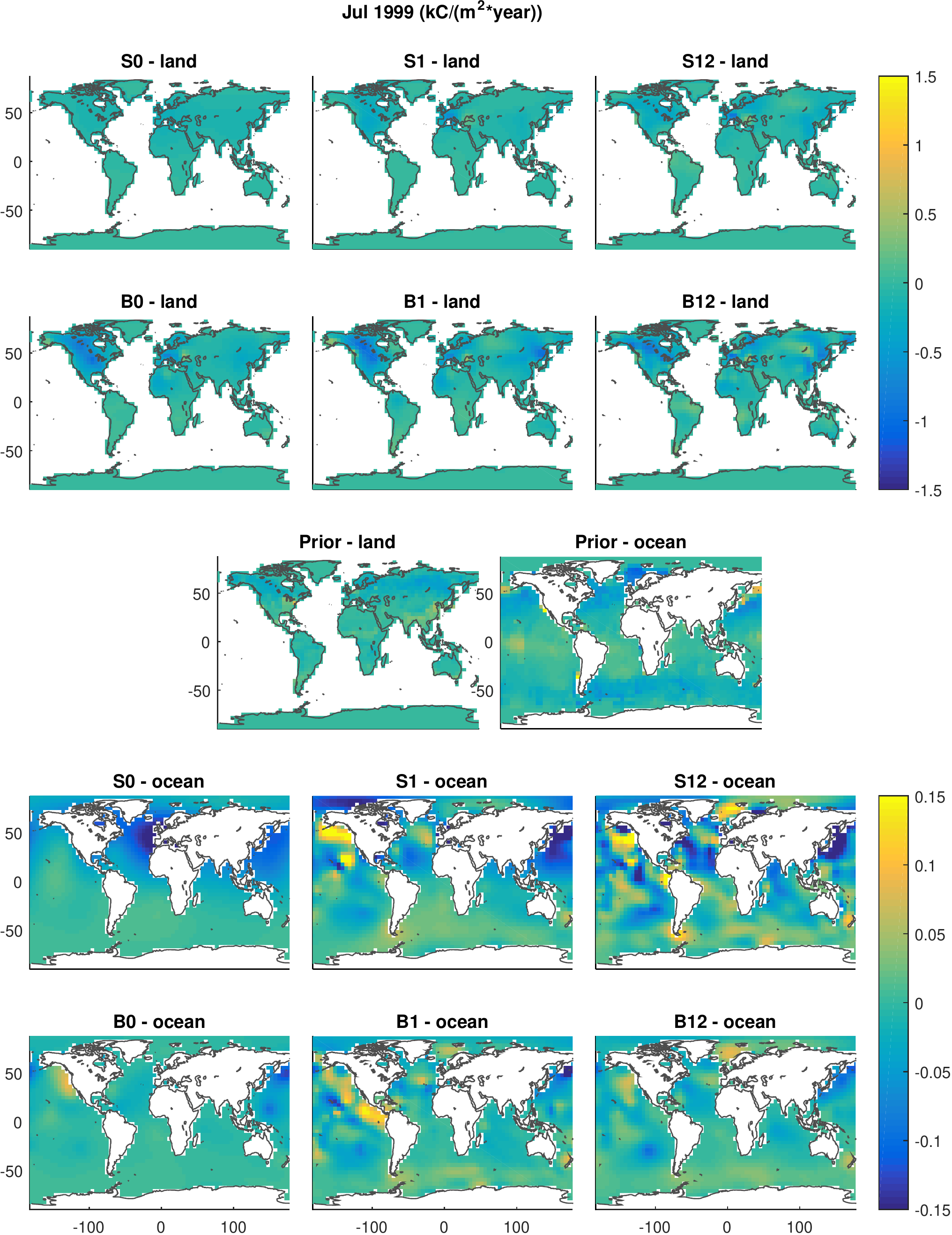}%
 \caption{Estimated flux anomalies for July 1999, using the six different models. The first two rows show the land flux anomalies, the third row shows the prior land and ocean fluxes, and the last two rows display the ocean anomalies. Note that the colour scale differs for land and ocean fluxes.}
   \label{fig:july}
\end{figure}

\begin{figure}[htb]
\begin{center}
   \includegraphics[width=0.99\linewidth]{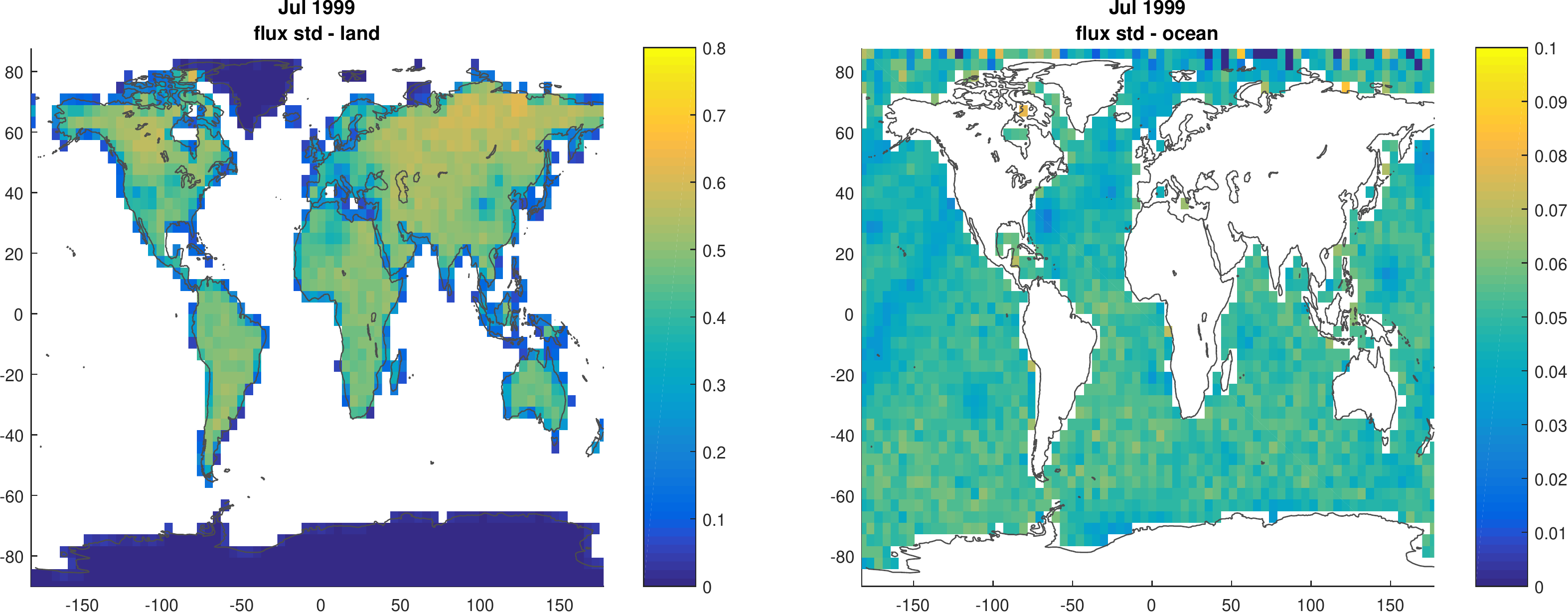}%
   \caption{Approximate standard deviations for the land and ocean anomalies estimated using the B12-model (the anomalies are presented in Fig.~\protect{\ref{fig:july}}) for July of 1999 [$\text{kgC}/(\text{year}\cdot \text{m}^2)$].}
   \label{fig:B12_rec1999_Jul}
  \end{center}
\end{figure}

Anomalies for a specific year and month (here July, 1999), estimated using the different models, are shown in Figure~\ref{fig:july} (anomalies for January, April and October can be found in the supplementary material). The anomalies were obtained by subtracting the prior flux, $\bm{\mu}_0$, from the posterior fluxes \eqref{eq:cond_fluxes_expectation}, and are important for determining which \CO fluxes that not captured by the prior fluxes. The effect from having separate models for land and ocean fields is clear; resulting in stronger signals over land and weaker signals over oceans. Moreover, having separate land and ocean fields allow us to better resolve fluxes on the Southern continents (South America and Africa). Figure~\ref{fig:B12_rec1999_Jul} illustrates the approximate posterior standard deviations for the B12-model during July of 1999 obtained using \eqref{eq:Bekas_f_simplified}. The posterior standard deviations are smaller for areas with many observational sites; e.g.\ in central Europe and at the west coast of North America. The same holds for the standard deviations of the ocean fluxes, with lower values along the measurement locations in the Pacific Ocean.

\clearpage

\section{Discussion}

\subsection{Model parameters}
\label{Model parameters}
It is interesting to compare the estimated parameters of our best model (Tbl.~\ref{tab:model_parameters}, model B12) with those found in other inversion studies. Firstly, our estimated marginal standard deviations for the land field is about ten times larger than for the ocean field. The higher variability of land fluxes is consistent with previous inversion studies \citep{mueller2008, bousquet2000}. 

For the spatial range most previous studies use an exponential covariance function $r(s) \propto \exp(-s/\ell)$ and report the parameter $\ell$ as the range. The link between $\ell$ and our definition of range is $\rho = \ell\cdot \log(10)$, and all ranges presented here have been transformed to follow our definition and are given in Earth radii. Additionally, previous studies either assume known spatial ranges \citep{rodenbeck2003} or estimate them from bottom-up fluxes \citep{mueller2008} and not directly from \CO observations as done here. Given these caveats, our spatial ranges are much smaller (see Table~\ref{tab:compare_range}), than those reported in previous studies. However, our final range estimates are consistent with those we obtained from the prior NEE and ocean fluxes using a single field (Bottom-up estimate, Sec.~\ref{sec:Simulation_study}).

\begin{table}[htb]
\caption{Comparison of our estimated spatial ranges for the latent flux field, first two lines, with those obtained or used in previous studies. Note that our bottom-up estimate is based on a single field for both land and ocean, and estimated from the prior NEE and ocean fluxes.}
\label{tab:compare_range} 
\begin{center}
\begin{tabular}{l|SS}
 & {Land} & {Ocean} \\ \hline
Bottom-up estimate & \multicolumn{2}{c}{0.118} \\
\CO observations & 0.0771 & 0.213 \\ \hline
\citet{rodenbeck2003} & 0.4605 & 0.6908 \\
\citet{mueller2008}   & 0.9758 & 2.0601 \\
\end{tabular}
\end{center}
\end{table}

Both \citet{rodenbeck2003} and \citet{michalak2004} use exponential covariances in time, equivalent to an AR(1), with parameters respectively assumed known or estimated from bottom-up fluxes. In contrast \citet{mueller2008} includes different intercepts for each month in the regression model; essentially creating a seasonal dependence, somewhat similar to our AR(12)-process with $a=0$ and $b=1$. In general, our inclusion of a seasonal AR(12)-process results in substantially larger temporal ranges, than in previous studies.

The strong seasonal component in our model can be motivated by the seasonal trend in the observed \CO concentrations that is not fully explained by the prior fluxes (Fig.~\ref{fig:val_trueData}). In contrast to the spatial range, the estimated temporal parameters (Tbl.~\ref{tab:model_parameters}) are much larger than those obtained for the prior NEE and ocean fluxes (Tbl.~\ref{tab:true_parameters}), indicating that the temporal dependence might be overestimated. The simulation study also had problems identifying the temporal parameters when using bottom-up flux fields. Theses issues might be due to non-Gaussian distributions of the latent fluxes and/or insufficient information in the inverse system to identify both temporal and spatial dependence. However, despite these problems, models using an AR(12) dependence consistently outperformed all other models both in-sample and for the validation data (Tbl.~\ref{tab:information_criterias}).

\subsection{Estimated Posterior fluxes}
\label{sec:estimated_fluxes}
\begin{figure}[htb]
\centering
   \includegraphics[width=0.8\linewidth]{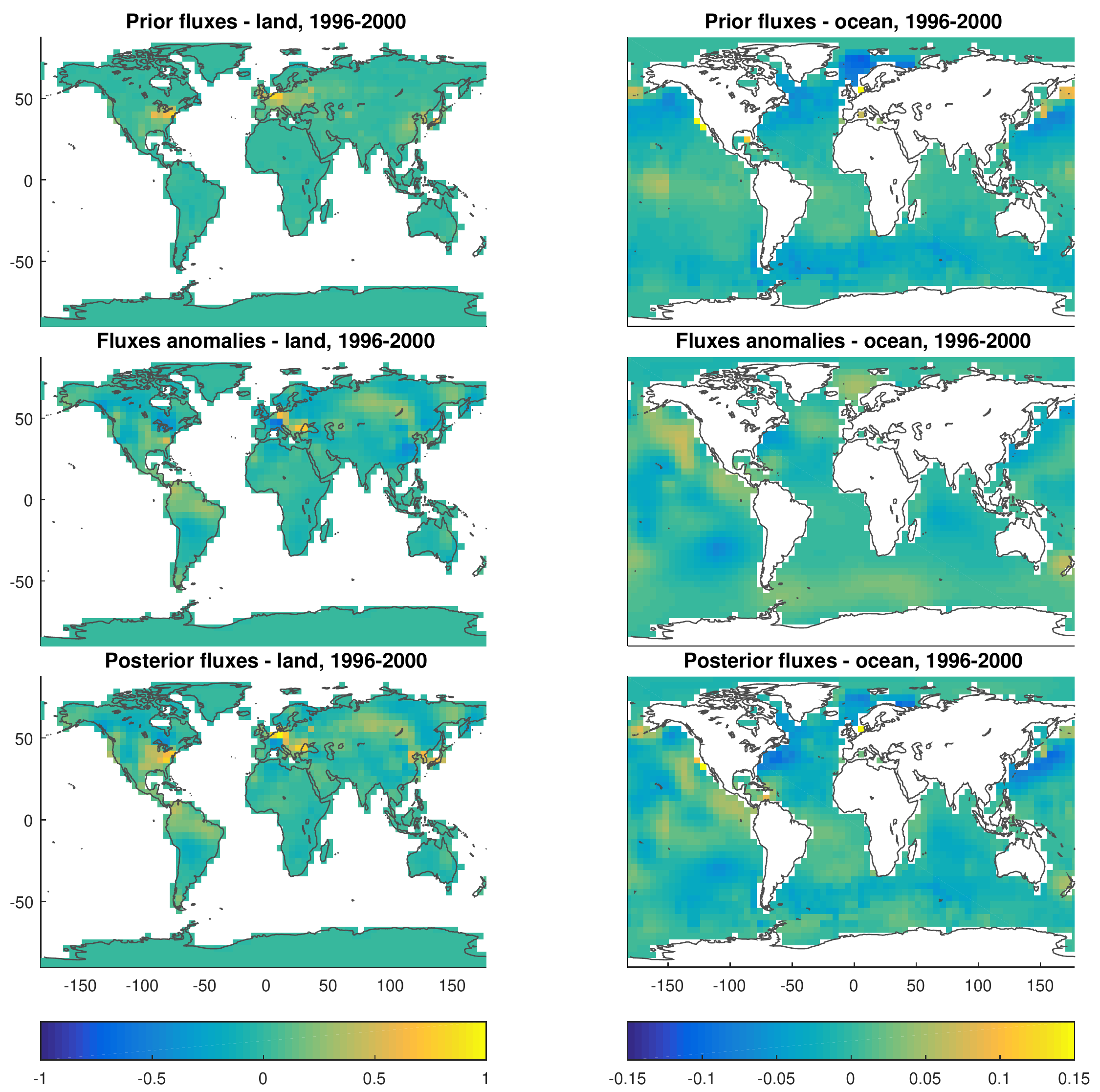}%
  \caption{Averaged land and ocean fluxes [$\text{kgC}/(\text{year}\cdot \text{m}^2)$] over the period 01/1996-12/2000 using the B12-model.}
   \label{fig:B12average}
\end{figure}

The reconstructed fluxes averaged across 1996--2000 and divided into
prior fluxes, spatial field (anomalies) and posterior fluxes are
shown in Figure~\ref{fig:B12average}. 
Two stronger sources in the
anomalies are apparent in Northern Germany and South-east Europe (west
of the black sea). We suspect, at least for South-east Europe, that this is due to
local pollution events from fossil fuel emissions close to the
measurement station in this region.
In tropical South America we
find a  dipole source/sink character with the northern part
of tropical South America being a source and southern a slight sink of
CO$_2$. The other tropical regions in Africa and Asia are close to
neutral or a small source. Note that the tropics, especially in South
America, are not well constrained by the observational data due to
the location of stations, Fig.~\ref{fig:observation_location}, and
substantial uncertainties in the reconstructions,
Fig.\ref{fig:B12_rec1999_Jul}. 

In the ocean anomalies, there is a recurrent
sink visible in the South Pacific and sources in the North Pacific and
in the Norwegian Sea. These patterns are obtained regardless of which
model is used to reconstruct fluxes. Remaining sources and sinks have
lower amplitudes and are more diffuse. 

\begin{figure}[htb]
\centering
   \includegraphics[width=0.8\linewidth]{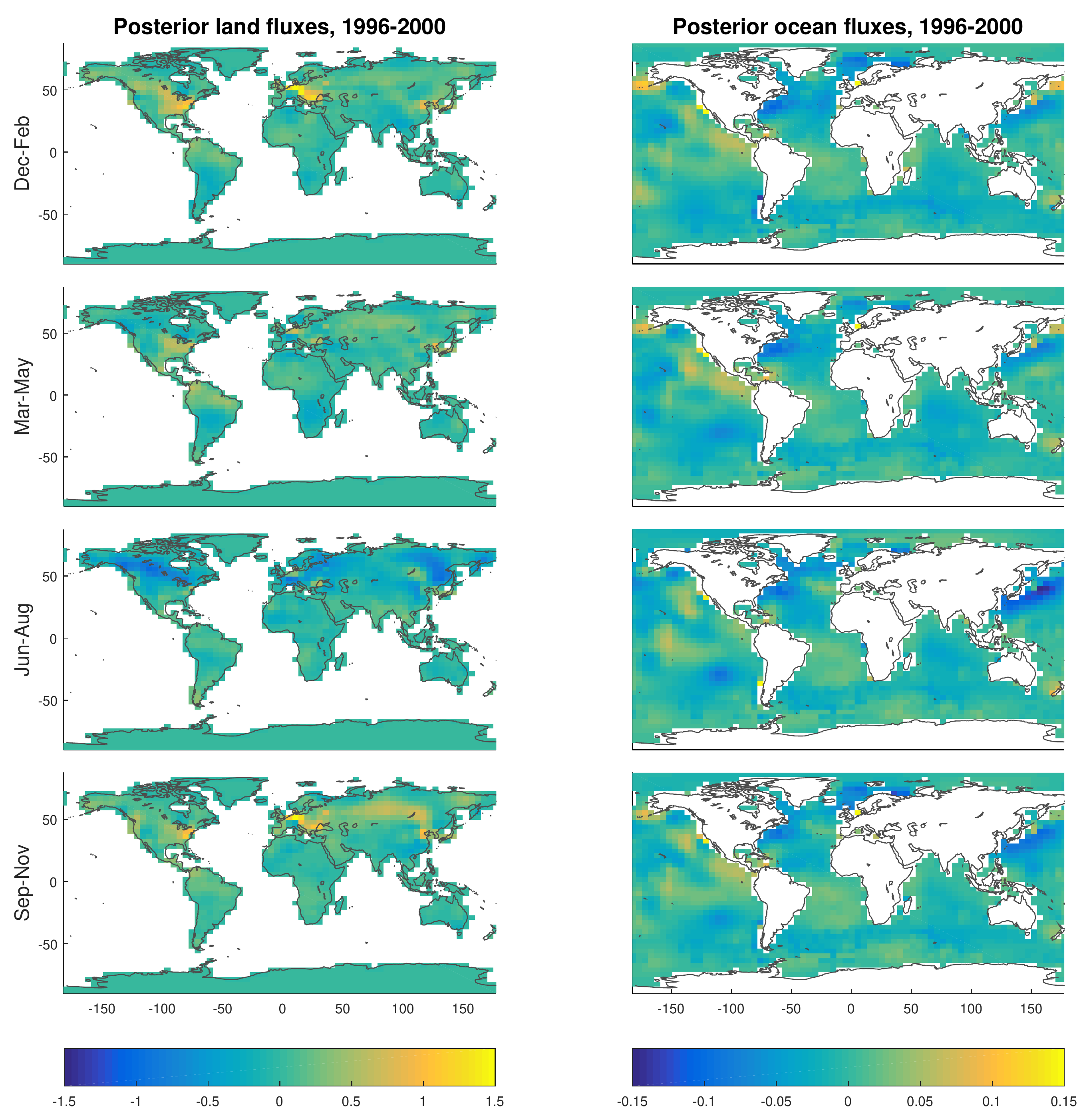}%
  \caption{Average seasonal land and ocean fluxes [$\text{kgC}/(\text{year}\cdot \text{m}^2)$] over the period 01/1996-12/2000 using the B12-model.}
   \label{fig:B12averageSeasons}
\end{figure}

Average seasonal fluxes are given in
Figure~\ref{fig:B12averageSeasons}. Here we can see the large sources in the
anomalies in central Europe (mentioned above) but also in the South Eastern USA are
mainly apparent in the Northern Hemisphere autumn and winter, i.e. at times when
the ecosystem respiration is dominating the terrestrial exchange
fluxes.  The largest sink occurs in North America and Eurasia during
the  Northern Hemisphere summer (growing season of
the vegetation) and caused by the uptake of \CO by the vegetation
through photosynthesis. For the ocean, there are substantial sinks
with a strong seasonality in the North Atlantic and east of Japan in the Pacific.

\subsection{Time series}
\label{sec:time_series}
\begin{figure}[!]
\begin{center}
   \includegraphics[width=0.9\linewidth]{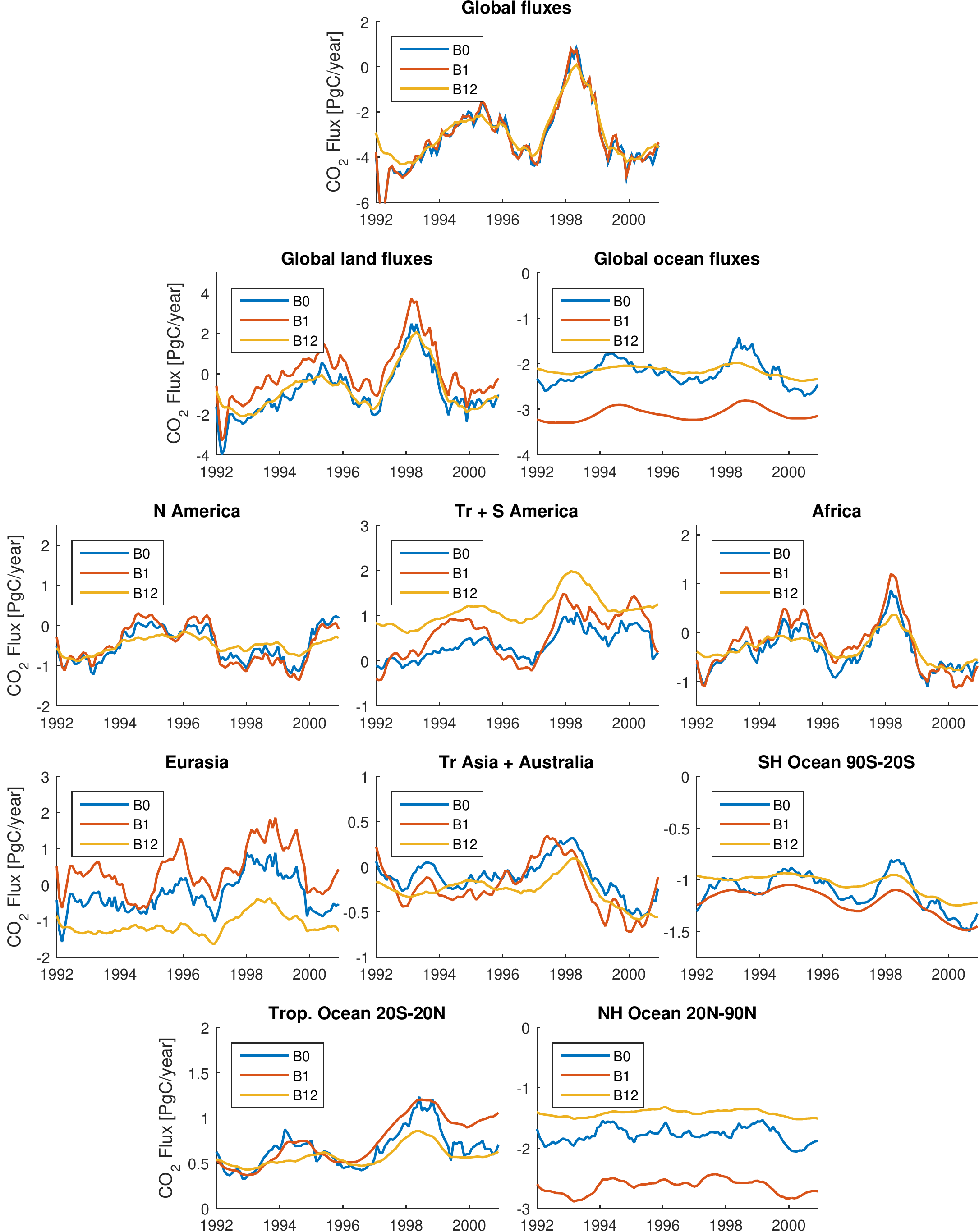}%
  \caption{Posterior fluxes integrated over different regions and deseasonalized. The top row gives global time-series at a monthly resolution, second row gives the land and ocean contributions, and the remaining rows provide regional fluxes.}
   \label{fig:timeseries}
  \end{center}
\end{figure}

Time series of reconstructed \CO fluxes integrated globally and over different regions are shown in Figure~\ref{fig:timeseries}, for the three models with both land and ocean fields. The contribution due to the fossil component in the prior fluxes, $\bm{\mu}_0$, has been removed, and the trends have been deseasonalized by averaging over running yearly intervals.. The regions used correspond to major continental landmasses, tropic and extra-tropical oceans; a map is provided in the supplementary material. 

As can be expected, models with more complex temporal dependence, B1 and B12, exhibit smoother temporal trends. The global times-series quite closely resemble those found by \citet{rodenbeck2003}, with increases of \CO fluxes during 1994--1995 and 1998. When the response is split into land and ocean parts, the main difference with \citet{rodenbeck2003} is the more distinct ocean peak for 1994--1995 seen in our inversion.

Across the eight individual regions we obtain quite different fluxes,
depending on which model is used. The largest differences occur in
Tropical and South America, Eurasia, and the Northern Oceans. This
suggests that regional fluxes may not be well-constrained by the
available observations in these areas, as already mention before in
Section~\ref{sec:estimated_fluxes}. In fact, the number of stations in these
regions is lower than in Europe\footnote{While Europe is part of
Eurasia, most of the stations are in Western Europe with almost no
stations in Siberia.} or North America. The large differences
between models for Eurasia and the Northern Ocean are correlated,
e.g.~the B12-model has the largest sinks over Eurasia and the smallest
sinks for the Northern Ocean whereas the opposite is true for the
B1-model. We suspect that this is a consequence of the spatial
distribution of observations stations across the Northern hemisphere
combined with predominant easterly transport due to the dominating
westerly wind fields.

\section{Conclusion}
In this article we introduced a new method for inverse modelling of global \CO surface fluxes based on GMRFs. In contrast to previous inversion methods, the definition of GMRFs as a basis expansion \eqref{eq:basis_expansion} allows for a spatially continuous representation of the fluxes. The observations response to fluxes is obtained by performing numerical integration over the different flux components, resulting in linear transport matrices. There are three main advantages of this method: (1) The GMRF model on the latent field provides a flexible system for constructing complex spatial and temporal covariances, demonstrated here by the inclusion of seasonal dependences. (2) The flux model represented on a continuous domain better represents the true fluxes, and reduces aggregation errors when combining data at different resolutions \citep{MoragCMP2017_21a}. (3) Computational advantages obtained through the use of sparse matrices and the Kronecker structure induced by a separable spatio-temporal covariance structure \citep{rue2004,BakkaRFRBIKSL2018_10}. 

In contrast to previous inversion studies, we estimate all model parameters using observations of \CO concentrations. Moreover, we show, using a simulation study, that the estimated parameters from \CO concentrations yield the best flux reconstructions, even when the true parameters of the system are available.

The best flux model obtained in this study, model B12 (see Table~\ref{tab:models}) has separate fields for land and ocean, and a strong seasonal dependence between months, e.g.\ January to January. The spatial dependencies in the model is shorter than in previous inversion studies (see Section~\ref{Model parameters}). This might be an effect of the stronger temporal dependence combined with an inability of the model to correctly identify both temporal and spatial dependencies due to the limited number of observations.

The GMRF model for fluxes presented here can be extended to non-stationary
covariances \citep{bolin2011,IngebLS2014_8}. This could account for possible differences in correlation strength due to e.g.\ latitude (different vegetation, climate zones and dominating wind directions) and land/ocean interactions. However, more unknown parameters, would require more data to constrain, and we have therefore limited this study to stationary covariance models. With increasing measurements from satellites such as NASA's Orbiting Carbon Observatory-2 \citep{OCO-2} the extension to a non-stationary covariance models could be interesting.

\medskip
Additional information and supporting material for this article is available online at the journal's website.
  
\section*{Acknowledgements}
This research is part of three Swedish strategic research areas: ModElling the Regional and Global Earth system (MERGE), the e-science collaboration (eSSENCE), and Biodiversity and Ecosystems in a Changing Climate (BECC). Dahlen and Lindström have been funded by Swedish Research Council (Vetenskapsrådet) grant no 2012–5983. Dahlen received financial support from Royal Physiographic Society of Lund. We thank Christian Rödenbeck for providing the TM3 transport Jacobian 
matrices.

\clearpage
\bibliographystyle{apalike}
\bibliography{CO2_reference}{}

\clearpage
\begin{appendices}
\renewcommand{\thesection}{\Alph{section}}
\section{Additional material} 
\label{App:Additional material}

\begin{figure}[ht]
\centering  
   \includegraphics[width=0.95\linewidth]{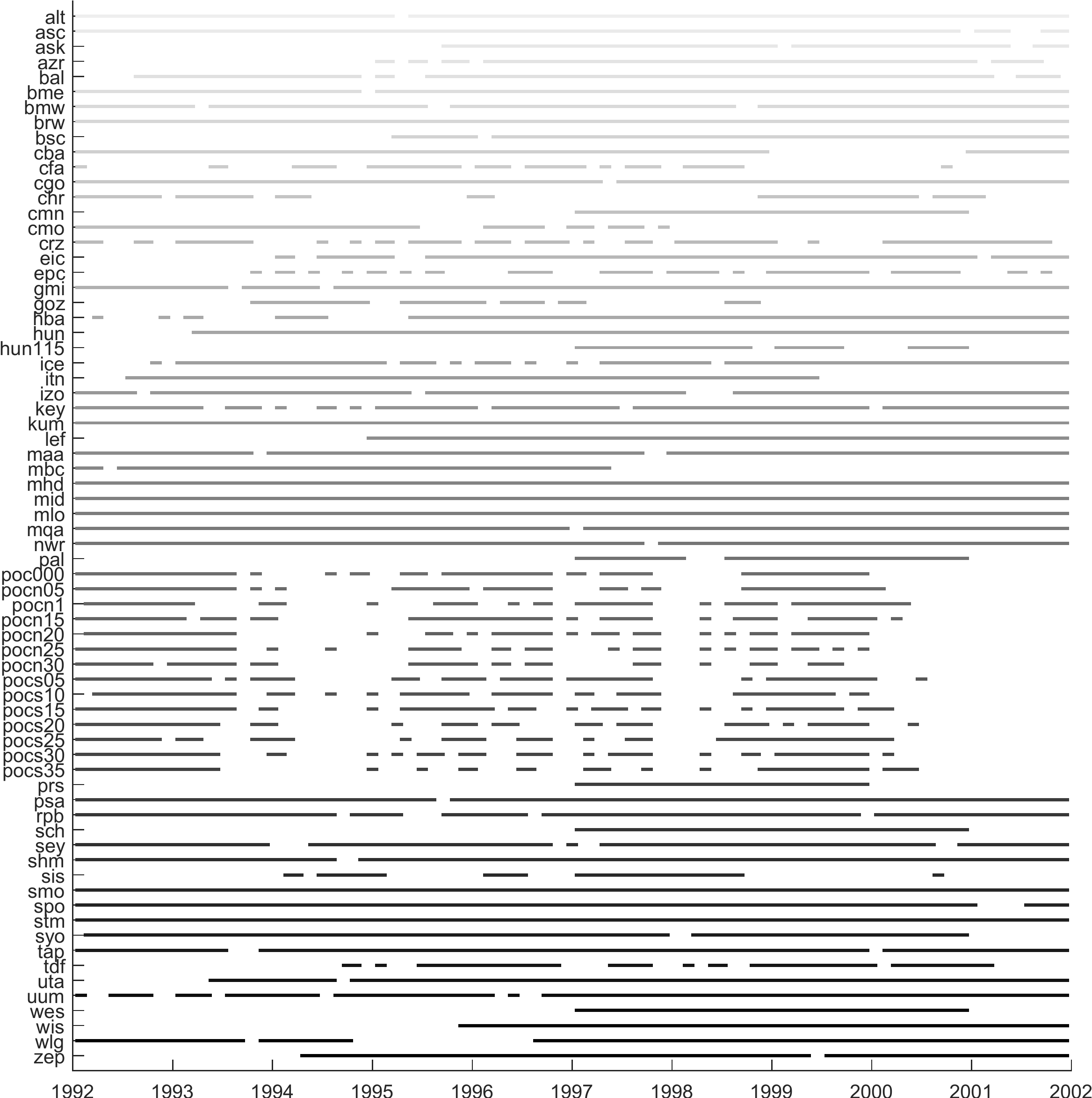}%
   \caption{The availability of observations over time for all stations. A more detailed description of the stations can be found in e.g. \citet{rodenbeck2005}.}
   \label{fig:observation_time}
\end{figure}

\clearpage

\section{Discretization} 
\label{App:Discretization}
To compute the relation between latent flux fields and observations we need to evaluate the integral over space and time in \eqref{eq:observation_model0}. Introducing an initial \CO concentration, $c_0$, and truncating the temporal integral reduces \eqref{eq:observation_model0} to \eqref{eq:observation_model1}. In practice the transport or sensitivity, $\bm{J}$, is defined at some spatial resolution (often a grid) on the globe, $\mathcal{J}=\{s_k\}_{k=1}^{n_s}$, and for regular time points $\{t_m\}_{m=1}^{n_t}$, where $n_s$ and $n_t$ are the spatial and temporal resolutions, respectively. Given a discrete representation of the sensitivity, i.e.\ $J(s_i,t_i,s_k,t_m)$, the spatially continuous observation model in \eqref{eq:observation_model1} can be written as
\begin{align*}
\begin{split}
y(s_i,t_i) &= 
c_0  + \sum_{m=1}^{n_t} \sum_{k=1}^{n_s} J(s_i,t_i,s_k,t_m) \int_{\bm{s}\in s_k} \!\! f(\bm{s},t_m) d\bm{s} + \epsilon(\bm{s},t) 
\\ &= 
c_0  + \sum_{m=1}^{n_t} \sum_{k=1}^{n_s} J(s_i,t_i,s_k,t_m) \int_{\bm{s}\in s_k} \!\! \biggl(\mu_0(\bm{s},t_m)+ \mu_\beta(\bm{s},t_m)+  \sum_{\ell=1}^{n_\ell} \phi_\ell(\bm{s})\omega_\ell(t) \biggr) d\bm{s} + \epsilon(\bm{s},t),
\end{split}
\end{align*} 
where $s_k$ should be interpreted as a grid cell. Here, we have allowed for a flux field consisting of a stochastic spatial field given by a basis expansion, as in \eqref{eq:basis_expansion}, as well as a mean given by a constant and a regression component, see \eqref{eq:flux_integration}. The flux components ($\mu_0(\bm{s},t)$, $\mu_\beta(\bm{s},t)$, and $x(\bm{s},t)$) might be provided at resolutions that differ from the transport resolution, in which case they need to be integrated over the spatial domain specified by $\mathcal{J}=\{s_k\}_{k=1}^{n_s}$. Numerical integration of the flux components can be performed by forming matrices, $\bm{L}_{0}$, $\bm{L}_{\beta}$ and $\bm{L}_{\omega}$, that map each component to the transport grid.

\subsection{Discretization of mean components}
Let, $\bm{\mu}_{0}^{(\mathcal{U})}(s_u,t_m)$, be a discrete spatial representation of the constant mean at time $t_m$, given at a spatial resolution of $\mathcal{U} = \{s_u\}_{u=1}^{n_u}$ (again $s_u$ is a grid cell). The corresponding constant mean defined on the same resolution as the transport grid is given by
\begin{equation*}
\bm{\mu}_{0}^{(\mathcal{J})}(s_k,t_m) = 
\!\!\! \int\limits_{\bm{s} \in s_k} \!\!\! \mu_0(\bm{s},t_m)\, d\bm{s} \approx
\sum_{u=1}^{n_u}  \bm{\mu}_{0}^{(\mathcal{U})}(s_u,t_m) 
\!\!\!\!\!\! \int\limits_{\bm{s} \in s_k \cap s_u} \!\!\!\!\!\! 1\, d\bm{s} = 
\sum_{u=1}^{n_u} D^{(\mathcal{U},\mathcal{J})}_{ku} 
\bm{\mu}_{0}^{(\mathcal{U})}(s_u,t_m),
\end{equation*}
where 
the spatial mapping at time $t_m$ is given by a matrix with elements equal to the area of overlap between grid cells in the two spatial resolutions:
\begin{equation}
\label{eq:linear_elements}
D^{(\mathcal{U},\mathcal{J})}_{ku} \triangleq  \lvert s_k \cap s_u \rvert.
\end{equation}
Since the spatial integration defined by \eqref{eq:linear_elements} is independent of time the mapping between the spatio-temporal components, $\bm{\mu}_0^{(\mathcal{U})}$ and $\bm{\mu}_{0}^{(\mathcal{J})}$, can be formed by a tensor product
\begin{align}
  \bm{L}_{0} =  \bm{I}_{n_t} \otimes \bm{D}^{(\mathcal{U},\mathcal{J})},
  \label{eq:L0_app}
\end{align}
where $\bm{I}_{n_t}$ is an identity matrix of size $n_t$. Similarly, the basis functions used in the regression mean, e.g.\ $B_j(\bm{s},t)$ in \eqref{eq:reg}, are mapped to the transport grid using
\begin{align}
  \bm{L}_{\beta} &=  \bm{I}_{n_t} \otimes \bm{D}^{(\mathcal{V},\mathcal{J})}
  & &\text{with elements} &
  D^{(\mathcal{V},\mathcal{J})}_{kv} &\triangleq  \lvert s_k \cap s_v \rvert,
\end{align}
where $\mathcal{V} = \{s_v\}_{v=1}^{n_v}$ is the resolution at which covariates are provided. If the constant mean and/or covariates are provided at the same resolution as the transport matrix --- i.e.\ if $\mathcal{U}$ and/or $\mathcal{V}$ are equal to $\mathcal{J}$ --- the corresponding matrices, $\bm{L}_{0}$ and/or $\bm{L}_{\beta}$, simplify to identity matrices of suitable size.

\subsection{Discretization of the stochastic field}
\begin{figure}[ht] 
\centering
 \includegraphics[width=0.4\linewidth]{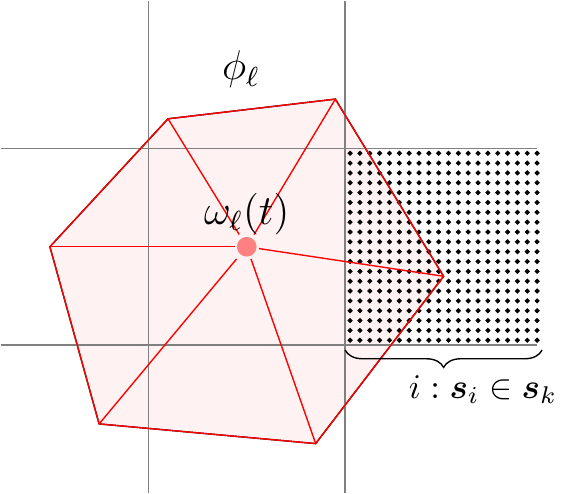}%
\caption{The numerical integration of basis functions to the observation grid, $s_k \in \mathcal{J}$, is done by evaluating the basis functions across a very dense grid, $s_i$, and replacing integrals by the corresponding approximate sums over $i$.}
\label{fig:numerical_integration}
\end{figure}

For the spatial field the integration over $\mathcal{J}$ is performed numerically. 
First we define a (very) dense grid with centre points $s_i$ (see Figure~\ref{fig:numerical_integration}) for each grid cell $s_k$ in $\mathcal{J}$. Given a basis expansion \eqref{eq:basis_expansion} of the spatial field, the basis functions are evaluated for each point in the dense grid, and the integrals are approximated using sums,
\begin{align}
\int_{\bm{s}\in s_k} \!\!\!\! x(\bm{s},t) d\bm{s}  &=
 \sum_{\ell=1}^{n_\ell} \left( 
   \int_{\bm{s}\in s_k} \!\!\!\!
   \phi_{\ell}(\bm{s})  d\bm{s} \right) \omega_{\ell}(t) 
 \approx
 \sum_{\ell=1}^{n_\ell} \left( 
 \sum_{\{i:s_i\in s_k\}} \!\!\! \phi_{\ell}(s_i)  \Delta s_i \right) 
 \omega_{\ell}(t).
 \label{eq:int_x}
\end{align}
Here $\Delta s_i$ represents the size of the grid cell centred at $s_i$; note that the grid cells will be of unequal size since the grid is defined on a sphere.

Introducing a matrix $\bm{G}$ with elements
\begin{equation}
G_{k\ell} \triangleq \sum_{\{i:s_i\in s_k\}} \!\!\! \phi_{\ell}(s_i) \Delta s_i,
\end{equation}
and following the same argument regarding repeated temporal fields as in \eqref{eq:L0_app}, the mapping from the weights in the spatio-temporal random field, $\bm{\omega}$, to the transport grid will be given by $\bm{L_\omega}=  \bm{I}_{n_t} \otimes \bm{G}$. 

Combining the spatio-temporal integration of all components, a discretized version of the transport model in \eqref{eq:observation_model1} can be written as,
\begin{align*}
\begin{split}
\bm{y} &=  c_{0} \cdot \bm{1} + \bm{J} \bm{L}_0\bm{\mu}_0 + 
  \bm{J} \bm{L}_\beta \bm{B} \bm{\beta} + 
  \bm{J} \bm{L}_\omega \bm{\omega} + \bm{\epsilon}
\\ &= 
  \underbrace{c_{0} \cdot \bm{1} + \bm{J} \bm{L}_0\bm{\mu}_0}_{\bm{y}_0} + 
  \bm{J} \underbrace{\begin{bmatrix}
  \bm{L}_\omega  & \bm{L}_\beta \bm{B} \end{bmatrix} }_{\bm{H}} 
  \underbrace{ \begin{bmatrix} \bm{\omega} \\ \bm{\beta}
  \end{bmatrix}}_{\bm{z}}  + \bm{\epsilon},
\end{split}
\end{align*}
which results in the observation model in \eqref{eq:flux_integration} and \eqref{eq:observation_model_Az}.

\section{Model Extension} 
\label{App:Model extension}
Different dynamics for continental and ocean fluxes can be obtained by using separate flux models for land and ocean. With the latent field taking the land flux value if $\bm{s}$ is over land and the ocean flux value otherwise, we obtain a spatio-temporal flux model given by
\begin{equation}
f(\bm{s},t)  = \mathbb{I}_{la}(\bm{s}) f_{la}(\bm{s},t)  + \mathbb{I}_{oc}(\bm{s}) f_{oc}(t, \bm{s}).
\end{equation}
where, $\mathbb{I}_{la}$ and $\mathbb{I}_{oc}$ are indicator functions defined as
\begin{align*}
\mathbb{I}_{la}(\bm{s}) &= \begin{cases}
    1, & \text{if $\bm{s}\in$ land,}\\
      0, & \text{if $\bm{s}\in$ ocean,}
  \end{cases}
  & 
  \mathbb{I}_{oc}(\bm{s}) &= \begin{cases}
    0, & \text{if $\bm{s}\in$ land,}\\
      1, & \text{if $\bm{s}\in$ ocean.}
  \end{cases}
\end{align*}
As before the individual fluxes, $f_{la}(\bm{s},t)$ and $f_{oc}(\bm{s},t)$, are represented on the form \eqref{eq:target_model}; now with separate mean models and spatio-temporal dependencies. The full flux model, cf.~\eqref{eq:flux_integration}, becomes
\begin{align}
\label{eq:flux_model_extended}
\bm{f} & = \bm{L}_{0}\bm{\mu}_{0} +  \underbrace{\begin{bmatrix} \bm{L}_{\omega_{la}}  & \bm{L}_{\omega_{oc}}  & \bm{L}_{\beta_{la}}\bm{B} &   \bm{L}_{\beta_{oc}}\bm{B} \end{bmatrix}}_{\bm{H}} 
\underbrace{\begin{bmatrix}
\bm{\omega}_{la} \\
\bm{\omega}_{oc}\\
\bm{\beta}_{la} \\
\bm{\beta}_{oc} 
\end{bmatrix}}_{\bm{z}},
\end{align}
where the $\bm{L}$ matrices are computed by accounting for the land/ocean indicators in the numerical integration detailed in Section~\ref{sec:observation_model} and Appendix~\ref{App:Discretization}. Assuming prior uncorrelated land and ocean fluxes, the distribution for the latent $\bm{z}$-field is
\begin{align}
\bm{z}=   \begin{bmatrix}
\bm{\omega}_{la} \\
\bm{\omega}_{oc}\\
\bm{\beta}_{la} \\
\bm{\beta}_{oc} 
\end{bmatrix}  &\in N \left( \begin{bmatrix} 
\bm{0} \\ \bm{0} \\ \bm{0} \\ \bm{0} 
\end{bmatrix}, \begin{bmatrix}
\bm{Q}_T(a_{la}) \otimes \bm{Q}_{S}(\bm{\theta}_{la}) & \bm{0} & \bm{0}  & \bm{0}  \\ 
\bm{0} & \bm{Q}_T(a_{oc}) \otimes \bm{Q}_{S}(\bm{\theta}_{oc})   &\bm{0} & \bm{0}  \\ 
\bm{0} & \bm{0} & \bm{Q}_{\beta_{la}} & \bm{0}  \\
 \bm{0} & \bm{0}  & \bm{0}  & \bm{Q}_{\beta_{oc}}
\end{bmatrix} \right).
\end{align}

\clearpage

\clearpage 
\section*{Supplementary Material} 
\begin{figure}[t]
\centering
   \includegraphics[width=0.7\linewidth]{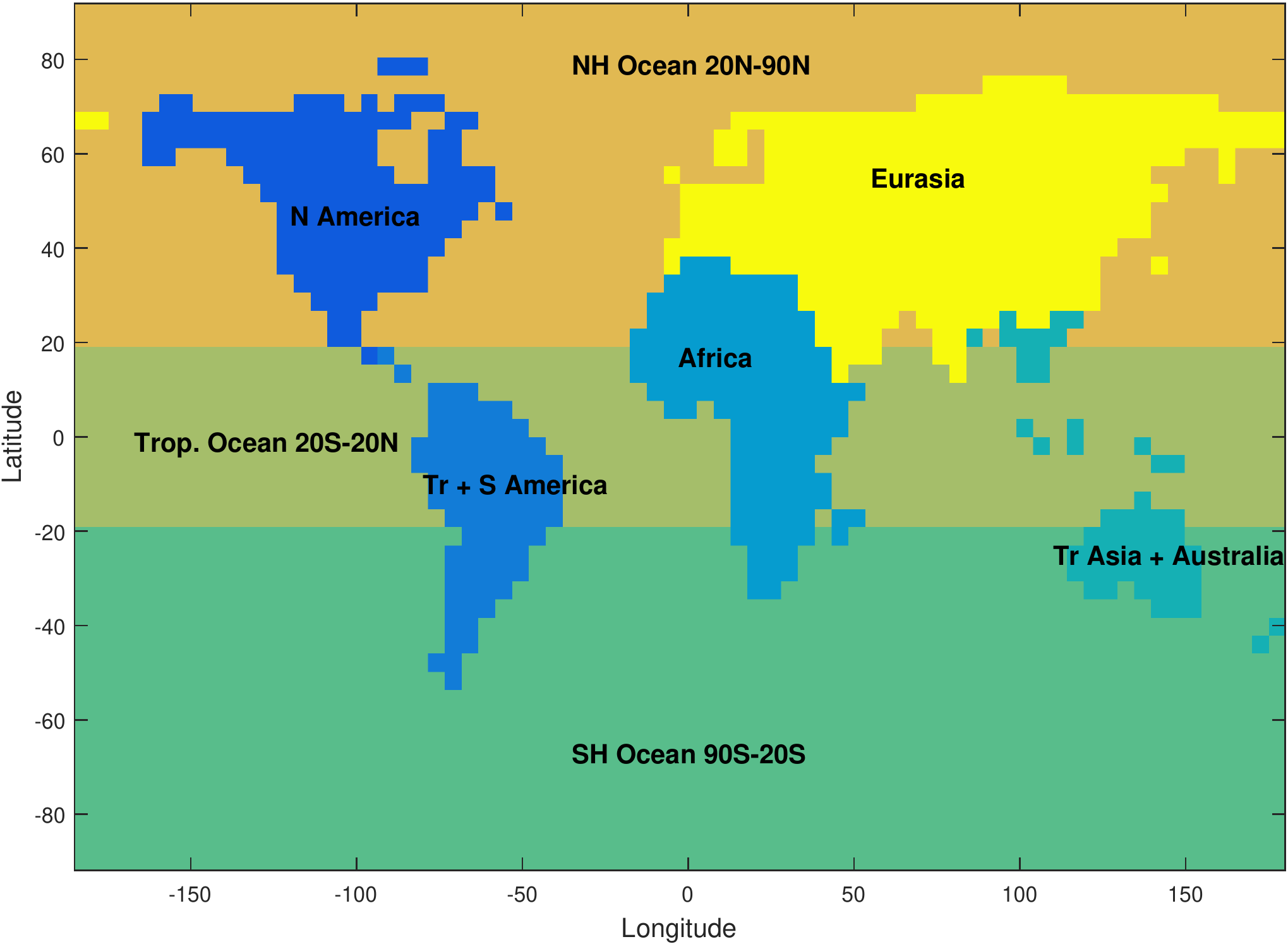}%
   \caption{Earth divided into eight regions, five continental regions and three ocean regions, for analysis of regional trends in \CO.}
   \label{fig:regions}
\end{figure}

\begin{figure}[!ht]
\centering
   \includegraphics[width=0.9\linewidth]{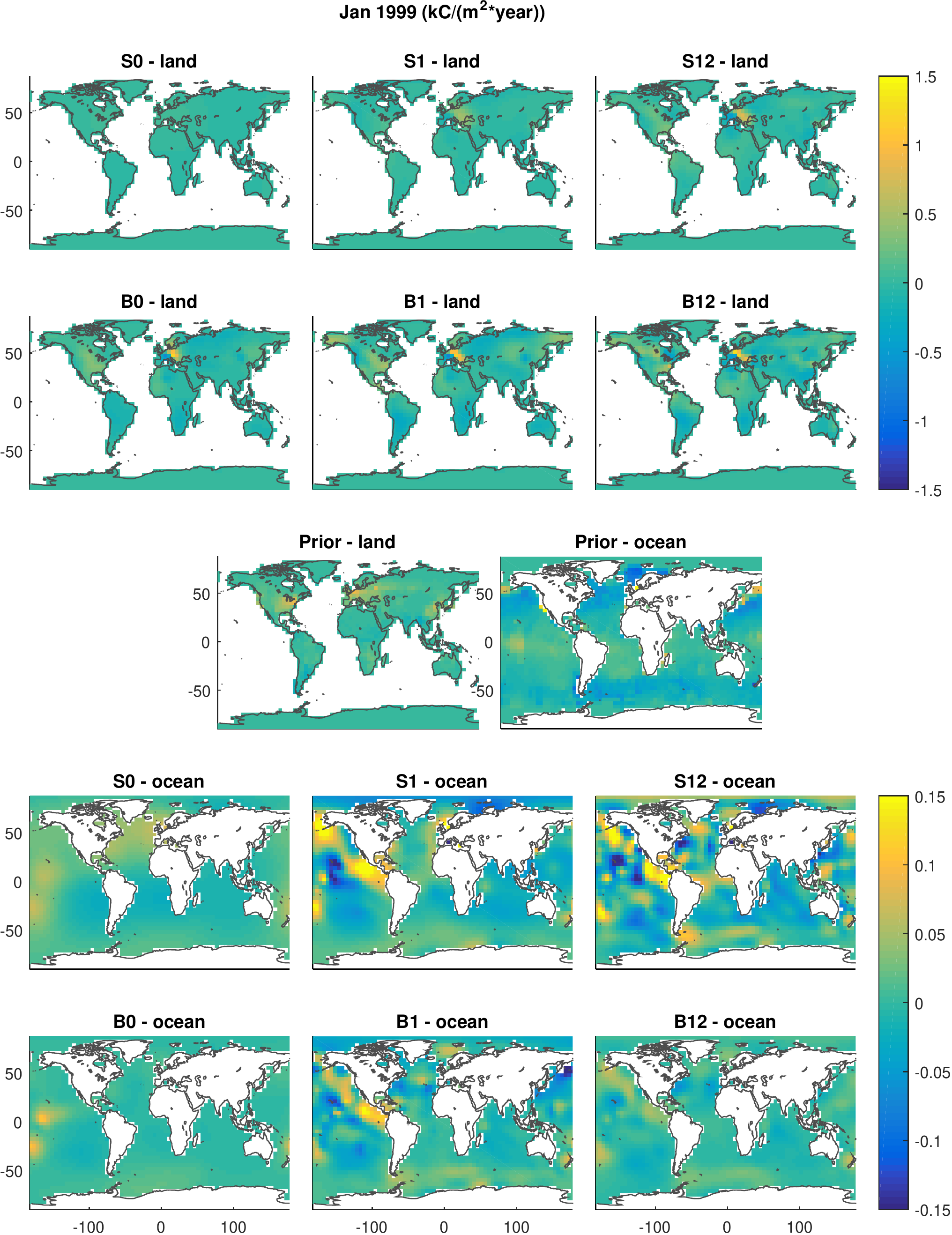}%
 \caption{Flux anomalies for January 1999, using the six different models. The first two rows show the estimated land flux anomalies, the third row shows the prior mean land and ocean fluxes, and the last two rows display the estimated ocean anomalies. Note that the colour scale differs for land and ocean fluxes.}
   \label{fig:january}
\end{figure}

\begin{figure}[!ht]
\centering
   \includegraphics[width=0.9\linewidth]{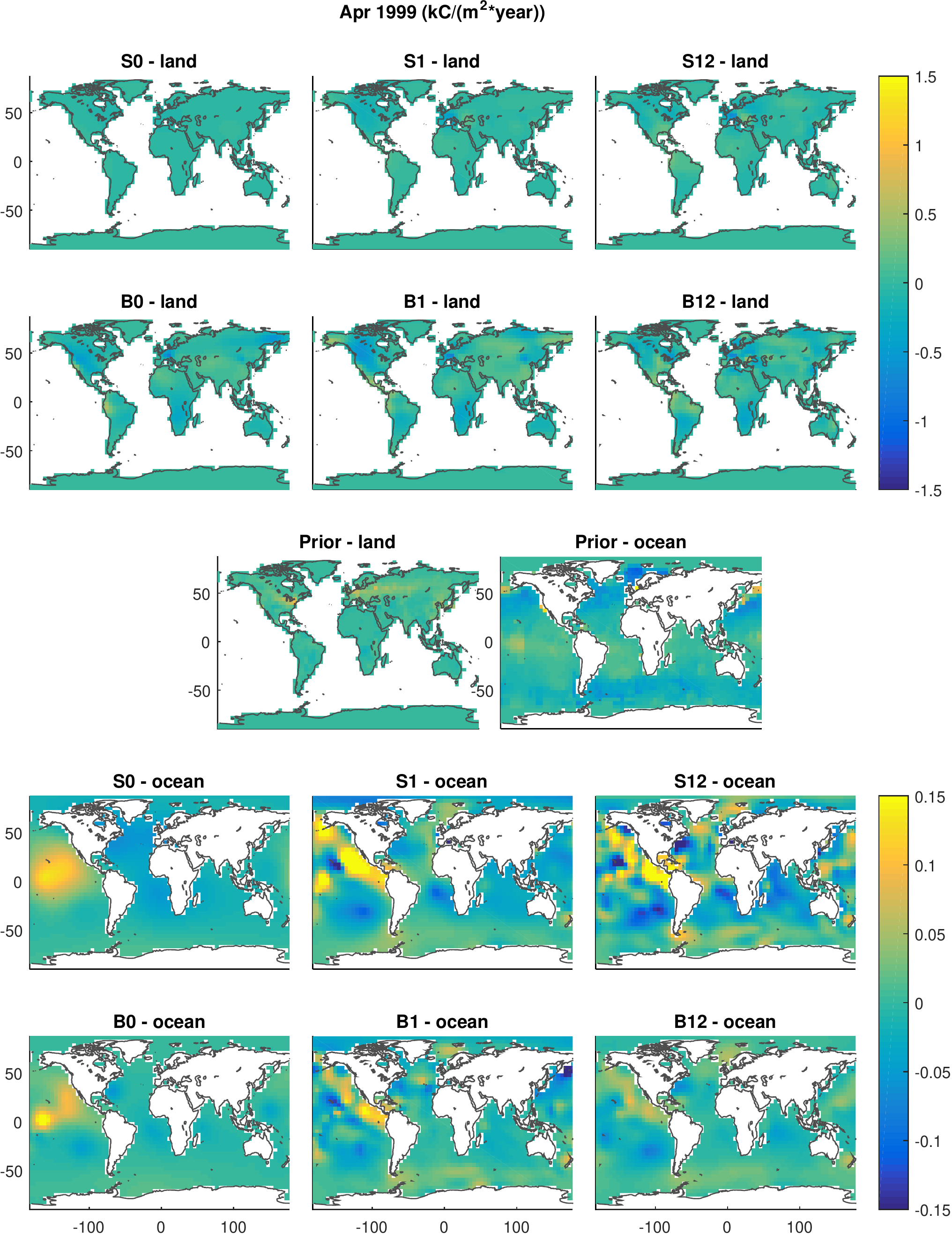}%
 \caption{Flux anomalies for April 1999, using the six different models. The first two rows show the estimated land flux anomalies, the third row shows the prior mean land and ocean fluxes, and the last two rows display the estimated ocean anomalies. Note that the colour scale differs for land and ocean fluxes.}
   \label{fig:april}
\end{figure}

\begin{figure}[!ht]
\centering
   \includegraphics[width=0.9\linewidth]{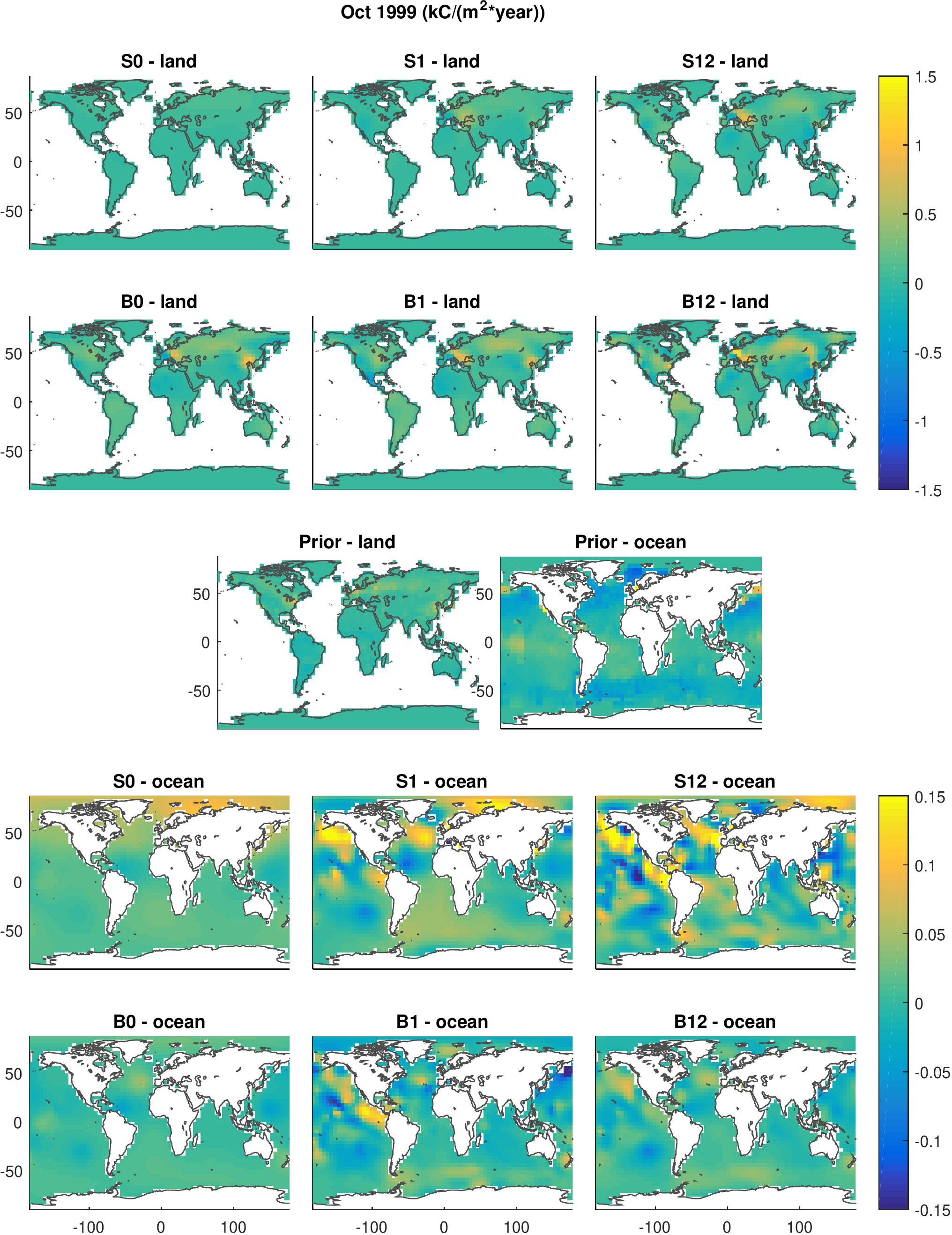}%
 \caption{Flux anomalies for October 1999, using the six different models. The first two rows show the estimated land flux anomalies, the third row shows the prior mean land and ocean fluxes, and the last two rows display the estimated ocean anomalies. Note that the colour scale differs for land and ocean fluxes.}
   \label{fig:october}
\end{figure}

\renewcommand{\thesection}{S\arabic{section}} 
\setcounter{section}{0} 
This supplementary material to the article \textit{Spatio-Temporal Reconstructions of Global CO2-Fluxes using Gaussian Markov Random Fields} provides some additional technical details and results not included in the main paper.
The precision matrix for a seasonal AR(12)-process and the Yule-Walker equations for computing corresponding covariance functions are provided in Section~\ref{App:Temporal precision matrix}. 
Section~\ref{App:Computational details} provides details of the likelihood simplifications using the \citet{woodbury1950} matrix identity and resulting computational complexities.
Finally, Section~\ref{App:Additional Results} provides additional figures and results. This includes:
The class specification and reconstruction error at the 70 measurement locations (Tbl.~\ref{tab:bias_rmse_class}); the division of earth into regions used in the analysis of regional trends (Fig.~\ref{fig:regions}); and flux anomalies for January, April and October of 1999 using all 6 models in Figures~\ref{fig:january}---\ref{fig:october}.

\section{Temporal precision matrix}
\label{App:Temporal precision matrix}
The temporal dependence in our model is obtained by driving an AR(1) or a seasonal AR(12) process with spatially dependent noise. The precision matrix for the resulting spatio-temporal process is given by the Kronecker product between the precision for the AR-process and the driving noise \citep[Ch.~7]{blangiardo2015}: $\bm{Q}_\omega = \bm{Q}_T \otimes \bm{Q}_{S}$.

To determine the temporal precision we consider a seasonal AR(p) process with
\begin{align}
\label{eq:ARp}
x_t &= a x_{t-1} + b x_{t-p} + \eta_t, &
\eta_t \sim N(0,\sigma^2).
\end{align}
The conditional distribution is given by
\begin{align}
[x_t|x_{-t}] &= [x_t|x_{t-1},x_{t-p}] \in N(ax_{t-1} +bx_{t-p}, \sigma_e^2),  
\qquad  t > p > 1,
\end{align}
where $x_{-t}$ contains all elements occurring before $t$, i.e.\ $x_{-t} = \{x_\tau: t > \tau \geq 1\}$. The distribution for the vector $\bm{x}=\begin{bmatrix}x_T & x_{T-1} & \ldots & x_1\end{bmatrix}$ can be written as a product of conditional distributions and an initial stationary component:
\begin{align}
\begin{split}
p(\bm{x}) =& \left(\prod_{t=p+1}^T p(x_t|x_{t-1},x_{t-p})\right) p(x_p,\ldots,x_1) 
\propto \\ \propto & 
\exp\left( -\frac{1}{\sigma^2} \sum_{t=p+1}^T {(x_t-ax_{t-1}-bx_{t-p})}^2 \right) 
p(x_p,\ldots,x_1) =
\\
=& \exp\left(- \frac{1}{\sigma^2}\bm{x}^T \bm{Q}_T \bm{x}\right),
\end{split}
\end{align}
where we want to identify the (scaled) precision matrix, $\bm{Q}_T$. The quadratic sum above expands to
\begin{align}
\begin{split}
 \sum_{t=p+1}^T {(x_t-ax_{t-1}-bx_{t-p})}^2 =& 
 \sum_{t=p+1}^T x_t^2 +
 \sum_{t=p}^{T-1} a^2 x_t^2 +
 \sum_{t=1}^{T-p} b^2 x_t^2 -
 \sum_{t=p+1}^T 2a x_t x_{t-1} + 
 \\ & +
 \sum_{t=p+1}^T 2b x_t x_{t-p}  +  
 \sum_{t=p+1}^T 2ab x_{t-1} x_{t-p}.
\end{split}
\label{eq:quadratic_QT}
\end{align}
Given a stationary initial distribution for $p(x_p,\ldots,x_1)$ the final elements of the time-series will be stationary, and by symmetry the top left and lower left corner of a stationary precision matrix $\bm{Q}_T$ have to be equal. Identifying the elements in \eqref{eq:quadratic_QT} with corresponding elements in the quadratic form $\bm{x}^T \bm{Q}_T \bm{x}$ the temporal precision matrix is:
\begin{equation*}
\bm{Q}_T  = \kbordermatrix{\mbox{indices}&1&2&3& \ldots &p&p+1 &p+2 &\ldots \\
1&1 & -a &  &  & 0 & -b & & \\
2&-a & 1+a^2 & -a &  & &  ab &-b & \\
3 & & -a & 1+a^2  & -a & &  & ab & \ddots & & \\
\vdots &  & &   & \ddots & & & &  \\
p & 0 & & & -a & 1+a^2 & -a& & \\
p+1 & -b & ab & &  & -a & 1+a^2+b^2 & & \\
p+2 & &  -b&  ab & &  &\ddots &  & \\
\vdots & & & \ddots & & & & & \\
}.
\end{equation*}

The covariance function, $r(k)=\mathsf{C}(x_t,x_{t-k})$, of the seasonal AR(p)-process can be found by solving the $(p+1)$-by-$(p+1)$ Yule-Walker equations \citep[Ch.~8]{Brock2009}:
\begin{align}
\begin{bmatrix}
1 & -a & 0 & & \ldots & & -b \\
-a & 1 &  0 & & \ldots & -b & 0 \\
& & & \ddots & & & \\
0 & -b & 0 & \ldots & -a & 1 & 0 \\
-b & 0 & & \ldots & & -a & 1 \\
\end{bmatrix}
\cdot
\begin{bmatrix}
r(0) \\ r(1) \\ \vdots \\ r(p-1) \\ r(p)
\end{bmatrix}
=
\begin{bmatrix}
\sigma_e^2 \\ 0 \\ \vdots \\ 0 \\ 0
\end{bmatrix},
\end{align}
and using the recursion
\begin{align}
r(k) &= a \cdot r(k-1) + b \cdot r(k-p),
\qquad k>p.
\end{align}
For a pure seasonal component, i.e.\ $a=0$ and $p>1$, we have
\begin{align}
r(k) &= \begin{cases} 
\sigma_e^2 \frac{b^{|l|}}{1-b^2}, & k=l p \qquad l\in\mathbb{Z} \\
0, & \text{otherwise,} \\
\end{cases}
\end{align}
and for the standard AR(1)-process ($b=0$):
\begin{align}
r(k) &= \sigma_e^2 \frac{a^{|k|}}{1-a^2}.
\end{align}

\section{Computational details} 
\label{App:Computational details}

Parameters estimates for the model (consisting of a latent Gaussian field with Gaussian observations) are obtained by maximising the log-likelihood \citep[see][for details]{rue2009}
\begin{equation}
  \label{eq:likelihood_sup}
   L_{\bm{\Psi}} \propto 
   \left( \frac{\lvert \bm{Q}_{\bm{z}} \rvert \lvert \bm{Q}_{{\epsilon}} \rvert}{\lvert \bm{Q}_{\bm{z}|\bm{y}} \rvert} \right) ^{1/2} 
   \exp \left( -1/2 \left[ 
     \bm{\mu}_{\bm{z}|\bm{y}}^T \bm{Q}_{\bm{z}} \bm{\mu}_{\bm{z}|\bm{y}} +
     (\bm{y} - \bm{y}_{0}  - \bm{\mu}_{\bm{z}|\bm{y}})^T \bm{Q}_{{\epsilon}} 
     (\bm{y} - \bm{y}_{0}  - \bm{\mu}_{\bm{z}|\bm{y}}) \right] \right). 
\end{equation}
Given parameter estimates reconstructions of the latent field are given by the conditional expectation
\begin{align}
\label{eq:mu_rec_sup}
\bm{\mu}_{{\bm{z}} | \bm{y}}(\bm{\Psi}) &= \bm{Q}_{\bm{z}|\bm{y}}^{-1}(\bm{\Psi}) \bm{A}^T\bm{Q}_{\epsilon}(\bm{\Psi})(\bm{y}-\bm{v_0}),
\end{align}
where the posterior precision is
\begin{align}
\label{eq:var_rec_sup}
\bm{Q}_{\bm{z}|\bm{y}}(\bm{\Psi})&= \bm{Q}_{\bm{z}}(\bm{\Psi})+\bm{A}^T\bm{Q}_{\epsilon}(\bm{\Psi})\bm{A}.
\end{align}

By the \citet{woodbury1950} matrix identity, the inverse of the posterior precision, $\bm{Q}_{{z|y}}$ \eqref{eq:var_rec_sup}, can be expressed as
\begin{equation}
\label{eq:Woodbury}
(\bm{Q}_z+ \bm{A}^T\bm{Q}_{\epsilon}\bm{A})^{-1} = \bm{Q}_z^{-1} - \bm{Q}_z^{-1}\bm{A}^T(\bm{Q}_{\epsilon}^{-1}+\bm{A}\bm{Q}_z^{-1}\bm{A}^T)^{-1}\bm{A}\bm{Q}_z^{-1}
\end{equation}
where $\bm{Q}_{\bm{z}}$ is a sparse precision matrix of size $[n_{z}\times n_{z}]$, $\bm{A}$ is a dense, due to the integrated observations, observation matrix of size $[n_{\text{obs}}\times n_{z}]$, and $\bm{Q}_{\epsilon}$ is a diagonal precision matrix of size $[n_{\text{obs}}\times n_{\text{obs}}]$. We will now show how \eqref{eq:Woodbury} can be used to simplify \eqref{eq:mu_rec_sup} and \eqref{eq:likelihood_sup}.

\subsection{Posterior mean} 
Inserting expression \eqref{eq:Woodbury} into the definition of the posterior mean \eqref{eq:mu_rec_sup}, we arrive at
\begin{eqnarray}
\bm{\mu_{z| y}} 
&=& \bm{Q}_z^{-1}\bm{A}^T\bm{Q}_{\epsilon}(\bm{y}-\bm{y}_{0})-(\bm{Q}_z^{-1}\bm{A}^T(\bm{Q}_{\epsilon}^{-1}+\bm{A}\bm{Q}_z^{-1}\bm{A}^T)^{-1}\bm{A}\bm{Q}_z^{-1})\bm{A}^T\bm{Q}_{\epsilon}(\bm{y}-\bm{y}_{0}). 
\label{eq:mu_woodbury_sup}
\end{eqnarray}
Replacing the precision, $\bm{Q}_z$, with its corresponding Cholesky decomposition; i.e.\ let $\bm{R}_{z}$ be an upper triangular matrix such that $\bm{R}_{z}^T \bm{R}_{z} = \bm{Q}_z$; we identify a repeating term, $\bm{S} = \bm{A}\bm{R}_{z}^{-1}$, and the posterior mean in \eqref{eq:mu_woodbury_sup} can be written as
\begin{eqnarray}
\bm{\mu_{z| y}} \label{eq:LAST_sup}
&=& \bm{R}_{z}^{-1}\bm{S}^T\bm{Q}_{\epsilon}(\bm{y}-\bm{y}_{0})-\bm{R}_{z}^{-1}\bm{S}^T(\bm{Q}_{\epsilon}^{-1}+\bm{S}\bm{S}^T)^{-1}\bm{S}\bm{S}^T\bm{Q}_{\epsilon}(\bm{y}-\bm{y}_{0}).
\end{eqnarray}

Since $\bm{R}_{z}$ inherits sparsity as well as block and Kronecker structure from $\bm{Q}_z$ we have 
\begin{align}
\bm{R}_{z} &= 
\begin{bmatrix}
\bm{R}_{\omega} & 0 \\
0 & \bm{R}_{\beta}
\end{bmatrix} =
\begin{bmatrix}
\bm{R}_{T}\otimes \bm{R}_{S} & 0 \\
0 & \bm{R}_{\beta}
\end{bmatrix},
\label{eq:R_sup}
\end{align}
where $\bm{R}_{[\cdot]}$ are the Choleskey factors of the corresponding $\bm{Q}_{[\cdot]}$-matrices. This allows for efficient computation of $\bm{S}$; see Section \ref{sec:linear system} below. Finally we introduce $\bm{L}$ as the following Choleskey factorization
\begin{align}
\bm{L}^T \bm{L} &= (\bm{Q}^{-1}_{\epsilon}+\bm{S}\bm{S}^T),
\label{eq:L_sup}
\end{align}
and $\bm{M} = \bm{S}^T (\bm{Q}_{\epsilon} (\bm{y}-\bm{y}_{0}))$. 
Plugging these intermediate computations into \eqref{eq:LAST_sup} the posterior mean in \eqref{eq:mu_rec_sup} becomes
\begin{equation}
\bm{\mu_{z | y}} = 
\bm{R}_{z}^{-1}( \bm{M}- \bm{S}^T \bm{L}^{-1} \bm{L}^{-T} \bm{S} \bm{M}).
\label{eq:mu_final_sup}
\end{equation}


\subsection{Likelihood}
In addition to the simplification of the posterior mean in \eqref{eq:mu_final_sup} we can also rewrite the determinant as \citep[Thm.~8.1]{Harvi1997},
\begin{equation}
\label{eq:det_sup}
\frac{\lvert \bm{Q}_{{\bm{z}}} \rvert \lvert \bm{Q}_{\epsilon} \rvert}{
	\lvert \bm{Q}_{\bm{z}|\bm{y}} \rvert} = 
\frac{1}{\lvert \bm{Q}_{\epsilon}^{-1} +  \bm{S}\bm{S}^T \rvert} = 
\frac{1}{\lvert\bm{L}\rvert^2}.
\end{equation}
The resulting simplified negative log likelihood is
\begin{align}
- \log L_{\bm{\Psi}}  & \propto  \log \lvert \bm{L} \rvert +  
\frac{1}{2} \left( \bm{\mu}_{{z|y}}^T \bm{Q}_{\bm{z}} \bm{\mu}_{{z|y}} + 
(\bm{y} - \bm{y}_{0}  - \bm{\mu}_{\bm{z}|\bm{y}})^T \bm{Q}_{{\epsilon}}
(\bm{y} - \bm{y}_{0}  - \bm{\mu}_{\bm{z}|\bm{y}})\right),
\label{eq:loglike_final_sup}
\end{align}
where $\bm{\mu}_{\bm{z}|\bm{y}}$ is computed using \eqref{eq:mu_final_sup}.

\subsection{Solving the linear system}
\label{sec:linear system}
The $\bm{S}$-matrix is computed by solving the triangular equation system $\bm{R}_{z}^T\bm{S}^T = \bm{A}^T$. Based on the block structure in $\bm{R}_z$, see \eqref{eq:R_sup}, the equation system can be separated as
\begin{align}
\bm{R}_{z}^T\bm{S}^T &= 
\begin{bmatrix}
\bm{R}_{\omega}^T & 0 \\
0 & \bm{R}_{\beta}^T
\end{bmatrix}
\begin{bmatrix}
\bm{S}_{\omega}^T \\
\bm{S}_{\beta}^T
\end{bmatrix} 
= 
\begin{bmatrix}
\bm{R}_{T}^T\otimes \bm{R}_{S}^T & 0 \\
0 & \bm{R}_{\beta}^T
\end{bmatrix}
\begin{bmatrix}
\bm{S}_{\omega}^T \\
\bm{S}_{\beta}^T
\end{bmatrix} 
= 
\begin{bmatrix}
\bm{A}_{\omega}^T \\
\bm{A}_{\beta}^T
\end{bmatrix}.
\end{align}
Thus, we need to solve two separate linear systems: $\bm{R}_{\beta}^T\bm{S}_{\beta}^T = \bm{A}_{\beta}^T$ and  $\bm{R}_{\omega}^T\bm{S}_{\omega}^T = \bm{A}_{\omega}^T$ (in case of both land and ocean flux fields, we obtain four linear systems), where the $\bm{R}$-matrices are sparse. The first system is $p$-dimensional, where $p$ is the number of covariates, since $p \ll n_\omega$ this will be much faster than solving the second system.

\begin{figure}[ht] 
\centering
 \includegraphics[width=0.7\linewidth]{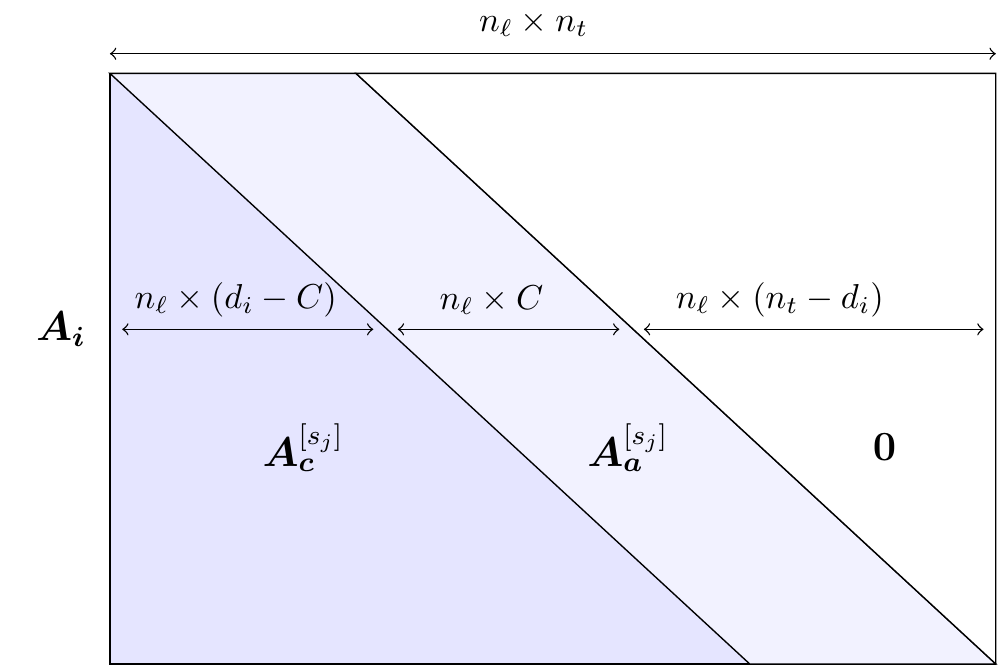} 
 \caption{Part of the observation matrix related to a single location, $s_j$;
	observations for this location are then ordered in time with later observations towards the bottom. The matrix consists of three main parts: 1) Left triangle  $\bm{A}_{c}^{[s_j]}$ --- The mapping for well mixed fluxes; 2) Band: $\bm{A}_{a}^{[s_j]}$ --- The mapping for the $C$ most recent flux fields; and 3) Right triangle: Zero-values indicating no sensitivity to future fluxes. A row in the matrix, as indicated by $\bm{A}_i$, gives the sensitivities for a single observation taken at time $d_i$.}
\label{fig:A}
\end{figure}
The cost for solving the latter equation system can be greatly reduced by making use of the Kronecker structure of $\bm{R}_{\omega}$ and the structure of the observation matrix, $\bm{A}_{\omega}=\bm{J}\bm{L_\omega}$. For the remainder of this section we simplify the notation by dropping the $\bm{\omega}$-subscripts. In Section~\ref{sec:Traditional inverse modeling} we noted that the transport matrix could be divided into two sub-matrices, $\bm{J} = \begin{bmatrix}
\bm{J}_{c} & \bm{J}_{a}
\end{bmatrix}$, with sensitivities to well mixed fluxes, and recent fluxes, respectively. 
The structure of the observation matrix at a single site, $s_j$, is illustrated in Figure~\ref{fig:A}. Here, $\bm{A}_{c}^{[s_j]}$ and $\bm{A}_{a}^{[s_j]}$ represents the non-zero elements of rows in $\bm{J}_{c}\bm{L_\omega}$ and $\bm{J}_{a}\bm{L_\omega}$, corresponding to location $s_j$. 

\begin{figure}[ht] 
\centering
 \includegraphics[width=0.6\linewidth]{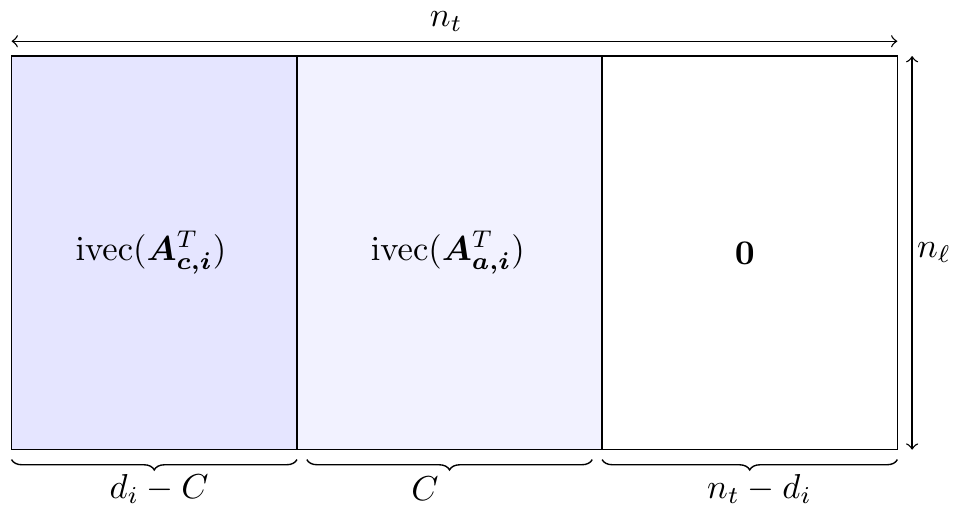} 
 \caption{The matrix $ivec(\bm{A}_i^T)$, composed into sub-matrices $ivec(\bm{A}_{{c,i}}^T)$, $ivec(\bm{A}_{{a,i}}^T)$, and a zero matrix.}
 \label{fig:ivecA}
\end{figure}

First, we note that,
\begin{equation}
\bm{R}^T\bm{S}^T = \bm{A}^T \Longleftrightarrow  \bm{R}^T\bm{S}_{i}^T = \bm{A}_{i}^T, \quad i=1,\ldots n_{\text{obs}}
\end{equation}
where $\bm{A}_{i}$ is the $i^\text{th}$ row in $\bm{A}$. Using the Kronecker structure of $\bm{R} = \bm{R}_T \otimes \bm{R}_S$ we have \citep{FernaPS1998_45},
\begin{align*}
\bm{R}^T\bm{S}_{i}^T = \bm{A}_{i}^T & \Leftrightarrow  
  \bm{S}_{i}^T = {(\bm{R}_T^T \otimes \bm{R}_S^T)}^{-1}\bm{A}_{i}^T   \\ 
& \Leftrightarrow    \bm{S}_{i}^T = (\bm{R}_T^{-T} \otimes \bm{R}_S^{-T})\bm{A}_{i}^T  \\ 
& \Leftrightarrow    \bm{S}_{i}^T = \text{vec} \left( \bm{R}_S^{-T} \text{ivec}(\bm{A}_{i}^T) \bm{R}_T^{-1}\right) \\ 
& \Leftrightarrow    \bm{S}_{i}^T = \text{vec} \left( \left(\bm{R}_T^{-T} (\bm{R}_S^{-T}\text{ivec}(\bm{A}_{i}^T))^T \right)^T \right),
\end{align*}
where $\text{ivec}(\bm{A}_{i}^T)$ is the column vector $\bm{A}_{i}^T$ of length $n_\ell  n_t$, reshaped to a matrix of size, $n_\ell\times n_t$, as illustrated in Figure~\ref{fig:ivecA}. The first $d_i-C$ columns in $\bm{A}_{i}$ are equal (sensitivities to well mixed fluxes), the next $C$ columns represent sensitivities to the $C$ flux fields just before observational time, and the following $n_t-d_i$ columns are zero (sensitivities to future fluxes). Moreover, the columns representing sensitivities to  well mixed fluxes are the same for all observations. As a result, $\bm{R}_S^{-T}\text{ivec}(\bm{A}_{i}^T)$ is achieved by essentially computing only the "recent" part $\bm{R}_S^{-T}\text{ivec}(\bm{A}_{{a,i}}^T)$ at a cost of $\mathcal{O} (C n_{\ell}\log n_{\ell})$. With a total cost of $\mathcal{O}(n_{\text{obs}} C n_{\ell} \log n_{\ell}  + n_{\ell} \log n_{\ell})$ across all observations, since $\bm{R}_S$ will have $\mathcal{O} (n_{\ell}\log n_{\ell})$ non-zero elements \citep[p.~51]{rue2004} and we can reuse the $\bm{R}_S^{-T}\text{ivec}(\bm{A}_{{c,i}}^T)$ computations for the constant part.

For the temporal component the sparse triangular systems has to be solved for all observations and locations resulting in a total cost of $\mathcal{O} (n_{\text{obs}} n_{t} n_\ell) = \mathcal{O} (n_{\text{obs}} n_\omega)$, which is the dominating factor when computing  $\bm{S} = \bm{A}\bm{R}_{z}^{-1}$.

\subsection{Computational costs}
Considering the necessary computations; the Choleskey factors $\bm{R}_t$,
$\bm{R}_s$, and $\bm{L}$ scale as $\mathcal{O}(n_t)$, $\mathcal{O}(n_s^{3/2})$ and $\mathcal{O}(n_{\text{obs}}^3)$, respectively. The different computational cost depends on the sparsity due to the temporal and spatial GMRFs \citep[Ch.~2.4]{rue2004}. Given the Choleskey factors, computation of the determinants is linear, since the determinant of a triangular matrix is computed as the product of the diagonal elemets: $\lvert \bm{L} \rvert = \prod_i \bm{L}_{ii}$. As described above the cost of computing $\bm{S} = \bm{A}\bm{R}_{z}^{-1}$, using back substitution and vectorization of the Kronecker product, is $\mathcal{O} (n_\omega n_{\text{obs}})$.

This makes the $\bm{S}\bm{S}^T$-product in \eqref{eq:L_sup} the most expensive calculation in \eqref{eq:mu_final_sup} and \eqref{eq:loglike_final_sup}, at a cost of $\mathcal{O} (n_{\text{obs}}^2 n)$ since we typically have $n>n_{\text{obs}}$.

\section{Additional figures and results} 
\label{App:Additional Results}

\begin{center}
\captionsetup{width=\textwidth}
\begin{longtable}{lccccc}
\caption{Observation class (last column) and reconstruction errors for all observations sites. The mean mismatch (bias) and RMSE were computed by comparing observed time-series of \CO from 01/1996 to 12/2000 with those obtained from the reconstructed fields using either the prior mean fluxes (a-prior) or conditional expectations from the spatio-temporal model (a-posteriori). Apart from the six validation locations marked in grey, the sites were used when computing the conditional expectations (i.e.\ they represent in-sample validation).} 
\label{tab:bias_rmse_class} \\
\hline
\multicolumn{1}{c}{\textbf{Location}} & \multicolumn{2}{c}{\textbf{Mean bias (ppm)}} & \multicolumn{2}{c}{\textbf{RMSE (ppm)}} & \multicolumn{1}{c}{\textbf{Class}} \\
 & \textbf{a-priori} & \textbf{a-posteriori} & \textbf{a-priori} & \textbf{a-posteriori}&  
 \\
\hline
\endfirsthead
\multicolumn{6}{c}%
{\tablename\ \thetable\ -- \textit{Continued from previous page}} \\
\hline
\multicolumn{1}{c}{\textbf{Location}} & \multicolumn{2}{c}{\textbf{Mean bias (ppm)}} & \multicolumn{2}{c}{\textbf{RMSE (ppm)}} & \multicolumn{1}{c}{\textbf{Class}} \\
 & \textbf{a-priori} & \textbf{a-posteriori} & \textbf{a-priori} & \textbf{a-posteriori}&  
 \\
\hline
\endhead
\hline \multicolumn{6}{r}{\textit{Continued on next page}} \\
\endfoot
\hline
\endlastfoot
\rowcolor{lightgray} alt & 2.385 & -0.193 & 3.673 & 0.927 & S \\ 
asc & 1.113 & 0.050 & 1.514 & 0.356 & R \\ 
ask & 2.298 & 0.012 & 2.965 & 0.237 & R \\ 
azr & 2.057 & 0.048 & 2.764 & 0.946 & R \\ 
bal & 1.240 & -0.120 & 4.087 & 2.123 & C \\ 
bme & 2.006 & 0.145 & 2.661 & 0.687 & R \\ 
bmw & 1.772 & 0.042 & 2.533 & 0.701 & R \\ 
brw & 2.111 & 0.089 & 3.661 & 0.361 & S \\ 
bsc & -3.471 & -0.279 & 6.985 & 2.254 & C \\ 
cba & 1.181 & -0.579 & 3.547 & 1.234 & S \\ 
cfa & 0.568 & 0.001 & 1.436 & 0.730 & S \\ 
cgo & 0.579 & -0.042 & 0.704 & 0.158 & S \\ 
\rowcolor{lightgray} chr & 1.046 & 0.624 & 1.412 & 0.980 & R \\ 
cmn & 3.531 & 0.875 & 5.451 & 2.409 & C \\ 
cmo & 1.829 & 0.147 & 3.966 & 1.275 & S \\ 
crz & 0.248 & -0.013 & 0.667 & 0.444 & R \\ 
eic & 1.620 & 0.234 & 1.788 & 0.628 & R \\ 
epc & 2.128 & 0.062 & 4.091 & 1.547 & S \\ 
gmi & 1.436 & 0.077 & 2.115 & 0.700 & R \\ 
goz & 2.393 & 0.164 & 3.058 & 1.225 & R \\ 
hba & 0.369 & 0.071 & 0.580 & 0.231 & R \\ 
hun & 0.035 & 0.049 & 7.409 & 2.534 & C \\ 
hun115 & 0.771 & -0.661 & 6.216 & 1.741 & C \\ 
ice & 2.248 & -0.009 & 3.621 & 0.597 & R \\ 
itn & 0.439 & -0.793 & 3.663 & 2.179 & C \\ 
izo & 1.897 & -0.109 & 2.652 & 0.549 & R \\ 
\rowcolor{lightgray} key & 1.138 & -0.652 & 2.551 & 2.214 & S \\ 
\rowcolor{lightgray} kum & 1.549 & -0.242 & 2.216 & 0.843 & R \\ 
lef & 2.365 & -0.382 & 5.017 & 2.793 & C \\ 
maa & 0.298 & 0.075 & 0.489 & 0.197 & R \\ 
mbc & 2.031 & -0.007 & 3.049 & 0.234 & S \\ 
\rowcolor{lightgray} mhd & 1.739 & -0.697 & 3.709 & 1.887 & S \\ 
mid & 1.595 & 0.074 & 2.310 & 0.593 & R \\ 
mlo & 1.811 & 0.195 & 2.211 & 0.472 & R \\  
mqa & 0.348 & 0.005 & 0.527 & 0.219 & R \\ 
nwr & 2.399 & 0.043 & 3.116 & 0.789 & M \\ 
pal & 2.301 & -0.077 & 4.433 & 0.945 & C \\ 
poc000 & 0.592 & -0.272 & 2.329 & 1.556 & R \\ 
pocn05 & 0.980 & 0.165 & 1.767 & 0.825 & R \\ 
pocn1 & 0.634 & -0.514 & 2.908 & 2.534 & R \\ 
pocn15 & 1.535 & 0.182 & 2.693 & 1.105 & R \\ 
pocn20 & 0.948 & -0.358 & 3.501 & 2.253 & R \\ 
pocn25 & 1.344 & -0.070 & 2.934 & 1.171 & R \\ 
pocn30 & 1.808 & -0.057 & 3.105 & 1.106 & R \\ 
pocs05 & 1.175 & 0.035 & 1.571 & 0.580 & R \\ 
pocs10 & 0.838 & -0.037 & 1.745 & 1.036 & R \\ 
pocs15 & 1.026 & 0.050 & 1.366 & 0.589 & R \\ 
pocs20 & 0.692 & 0.105 & 1.394 & 0.531 & R \\ 
pocs25 & 0.406 & 0.002 & 1.499 & 0.852 & R \\ 
pocs30 & 0.664 & 0.076 & 1.093 & 0.610 & R \\ 
pocs35 & 0.210 & -0.133 & 0.914 & 0.693 & R \\ 
prs & 3.055 & 0.737 & 4.088 & 1.470 & C \\ 
\rowcolor{lightgray} psa & -0.273 & -0.169 & 0.544 & 0.432 & R \\ 
rpb & 1.875 & 0.068 & 2.664 & 0.503 & R \\ 
sch & 4.905 & 0.054 & 5.879 & 0.501 & R \\ 
sey & 1.052 & -0.134 & 1.351 & 0.601 & R \\ 
shm & 3.167 & 0.087 & 5.111 & 0.550 & R \\ 
sis & 2.525 & -0.052 & 3.762 & 0.741 & R \\ 
smo & 1.319 & 0.012 & 1.455 & 0.313 & R \\ 
spo & 0.279 & -0.052 & 0.498 & 0.135 & R \\ 
stm & 1.772 & 0.032 & 3.078 & 0.469 & S \\ 
syo & -0.041 & -0.261 & 0.595 & 0.483 & R \\ 
tap & 0.722 & -0.164 & 3.816 & 2.197 & S \\ 
tdf & 0.089 & 0.075 & 0.740 & 0.429 & S \\ 
uta & 2.732 & 0.090 & 3.431 & 1.074 & C \\ 
uum & 2.335 & -0.008 & 3.797 & 1.147 & C \\ 
wes & 0.857 & -0.077 & 4.121 & 1.138 & S \\ 
wis & 1.731 & -0.013 & 3.154 & 0.968 & C \\ 
wlg & 2.976 & 0.004 & 3.907 & 0.313 & M \\ 
zep & 2.099 & -0.026 & 3.377 & 0.611 & R
\end{longtable}
\end{center}

\clearpage
\subsection{Flux anomalies for January, April, and October 1999}

\end{appendices}

\end{document}